\documentclass[aps,prl,reprint,superscriptaddress,longbibliography]{revtex4-1}
\usepackage{amsmath}
\usepackage{graphicx}
\usepackage{amsfonts}
\usepackage{color}
\usepackage{upgreek}
\usepackage{bm}
\usepackage{lipsum}
\usepackage{comment}
\usepackage{enumerate}
\usepackage{dsfont}

\usepackage[table]{xcolor}
\usepackage{textcomp}
\usepackage{amsmath}
\usepackage{amssymb}
\usepackage{graphicx} 
\usepackage{pgf}
\usepackage{epstopdf}
\usepackage{braket}
\usepackage{mathtools}
\usepackage{times}

\usepackage[retainorgcmds]{IEEEtrantools}
\newcommand{\bea}{\begin{eqnarray}}
\newcommand{\eea}{\end{eqnarray}}

\newcommand{\Tr}{\mathrm{Tr}}

\usepackage[colorlinks,linkcolor=blue,urlcolor=blue,citecolor=blue]{hyperref}

\begin{document}


\title{Bath-engineering magnetic order in quantum spin chains: An analytic mapping approach}



\author{Brett Min}
\email{brett.min@mail.utoronto.ca}
\affiliation{Department of Physics and Centre for Quantum Information and Quantum Control, University of Toronto, 60 Saint George St., Toronto, Ontario, M5S 1A7, Canada}

\author{Nicholas Anto-Sztrikacs}
\affiliation{Department of Physics and Centre for Quantum Information and Quantum Control, University of Toronto, 60 Saint George St., Toronto, Ontario, M5S 1A7, Canada}

\author{Marlon Brenes}
\affiliation{Department of Physics and Centre for Quantum Information and Quantum Control, University of Toronto, 60 Saint George St., Toronto, Ontario, M5S 1A7, Canada}

\author{Dvira Segal}
\email{dvira.segal@utoronto.ca}
\affiliation{Department of Chemistry,
University of Toronto, 80 Saint George St., Toronto, Ontario, M5S 3H6, Canada}

\affiliation{Department of Physics and Centre for Quantum Information and Quantum Control, University of Toronto, 60 Saint George St., Toronto, Ontario, M5S 1A7, Canada}

\begin{abstract}
Dissipative processes can drive different magnetic orders in quantum spin chains. Using a non-perturbative analytic mapping framework, we systematically show how to structure different magnetic orders in spin systems by controlling the locality of the attached baths. 
Our mapping approach reveals analytically the impact of spin-bath couplings,
leading to the suppression of spin splittings,
bath-dressing and mixing of spin-spin interactions, and 
 emergence of non-local {\it ferromagnetic} interactions between spins coupled to the same bath, which become long-ranged for a global bath. 
Our general mapping method can be readily applied to a variety of spin models: We demonstrate (i)
a bath-induced transition from 
antiferromangnetic (AFM) to ferromagnetic ordering in a Heisenberg spin chain, (ii)
AFM to extended Neel phase ordering
within a transverse-field Ising chain with pairwise couplings to baths, and (iii) a quantum phase transition in the fully-connected Ising model. 
Our method is non-perturbative in the system-bath coupling. It holds for a variety of non-Markovian baths and it can be readily applied 
towards studying bath-engineered phases in frustrated or topological materials. 
\end{abstract}\maketitle

\date{\today}

\textit{Introduction.---}
Spin chains offer a versatile platform for the study of quantum materials.
They can capture a wide range of complex and exotic phenomena 
from magnetic effects to topological phases. These effects are observed in a variety of materials, including quantum magnets, spin liquids, and quantum wires.
Beyond ideal models, in reality, environmental degrees of freedom such as lattice phonons or engineered cavity modes couple to the spin degrees of freedom. The resulting decoherence and dissipative effects may largely impact magnetic ordering in spin systems, even inducing quantum phase transitions, 
effects that stem from the interplay between internal spin-spin interactions and dissipation \cite{Winter_2014, De_Filippis_2021, Butcher_2022, Weber_2023,Landa_2020, Huybrechts_2020,
Orth_2008, Wang_2008, McCutcheon_2010,
Orth_2010, Grigoryan_2019,
Iemini_2018, Hannukainen_2018, Weisbrich_2018, Song_2023, Qian_2022, Garst_2004, Schehr_2008,
Nagele_2008, Hoyos_2012,
Sperstad_2012,  Bonart_2013,
Cai_2014, Goldstein_2015,
Maghrebi_2016,  Rose_2016,  Takada_2016,  Yan_2018,   Puel_2019, 
Landa_2020,  Huybrechts_2020,      
De_Filippis_2021,  Butcher_2022,  Weber_2022, Majumdar_2023, 
Perroni_2023,  Song_2023, Qian_2022, Lee_2013,  Chan_2015, Weimer_2015,  Jin_2016,  
Nakagawa_2020, Li_Jin_2021,
Overbeck_2017,  Rota_2017, 
Owen_2018,  Jin_2018,   Huber_2020,
Jaschke_2019, Lee_2011,  Li_2021}. 
These demonstrations were done using numerical approaches, facilitated by analytical arguments. 
The behavior of a collection of spins 
coupled to a {\it common} (global) bosonic bath was studied in 
Refs~\cite{Winter_2014, De_Filippis_2021,Butcher_2022,Weber_2023,Orth_2008,Wang_2008,McCutcheon_2010,Orth_2010,Iemini_2018,Hannukainen_2018,Weisbrich_2018,Grigoryan_2019,Landa_2020,Huybrechts_2020,Qian_2022,Song_2023}
, where it was demonstrated, using techniques such as the numerically-exact quantum Monte Carlo method \cite{Winter_2014, De_Filippis_2021,Butcher_2022,Weber_2023}, that such models can exhibit 
dissipation-controlled quantum phase transition. 
Other numerical studies focused on chains with sites {\it independently} (locally) coupled to 
dissipative baths \cite{Garst_2004, Cugliandolo_2005, Werner_2005, Werner_2_2005,Schehr_2008,Nagele_2008,Hoyos_2012,Sperstad_2012,Lee_2013,Bonart_2013,Cai_2014,Goldstein_2015,Chan_2015, Weimer_2015, Jin_2016, Maghrebi_2016, Rose_2016, Takada_2016, Overbeck_2017, Rota_2017, Owen_2018, Jin_2018,Yan_2018,Puel_2019, Landa_2020,Huybrechts_2020,Huber_2020,Nakagawa_2020, Li_Jin_2021,Weber_2022,Majumdar_2023,Perroni_2023,Weber_2023}  
demonstrating, e.g., long-range antiferromagnetic order at any coupling to the baths in an antiferromagnetic quantum Heisenberg chain.
%
Alternatively, other studies were done in the premise of weak system-bath couplings and/or structureless dissipation
using, e.g., the Lindblad quantum master equation \cite{Lee_2011,Lee_2013,Chan_2015,Weimer_2015,Jin_2016,Overbeck_2017,Rota_2017,Owen_2018,Jin_2018,Iemini_2018,Hannukainen_2018,Weisbrich_2018,Jaschke_2019,Landa_2020,Huybrechts_2020,Huber_2020,Nakagawa_2020,Li_Jin_2021,Song_2023}. 
While numerical studies of bath-controlled spin phases were often accompanied by analytical arguments; a rigorous, unified, and general analytic framework to bath-controlled phases is still missing.


Here, we
show that
a general mapping approach can be used to study a broad class of open spin systems
and provide an intuitive, {\it unified}, and comprehensive understanding of bath-induced phase transitions. 
Our method can treat global, local, or partially-local spin-bath coupling schemes at finite temperature and different families of baths' spectral density functions. 
The method is non-perturbative in the spin-bath coupling, thus enabling the capture of effects emerging from strong-coupling many-body physics. 
In a nutshell, 
based on unitary transformations and a controlled truncation, 
the mapping turns the spin+baths Hamiltonian into an {\it effective} Hamiltonian with the spin system now {\it weakly} coupled to its environments. 
Most crucially, our mapping approach reveals the generation of bath-mediated spin-spin interactions, which extend beyond nearest-neighboring spins---depending on the nonlocality of the attached baths, and favors ferromagnetic order. 
Bath-induced effects further mix and dress the intrinsic spin-spin couplings and suppress spin splittings. 
Through these bath-induced effects, our {\it closed-form}, Hermitian, effective spin Hamiltonian immediately evinces on the expected magnetic order as one tunes the system's couplings to its surroundings. 
We find that our mapping becomes more accurate as the size of the system is increased irrespective of the temperature of the bath.

\begin{figure}[htbp]
\fontsize{6}{10}\selectfont
\centering
\vspace{-9mm}
\includegraphics[width=1\columnwidth]{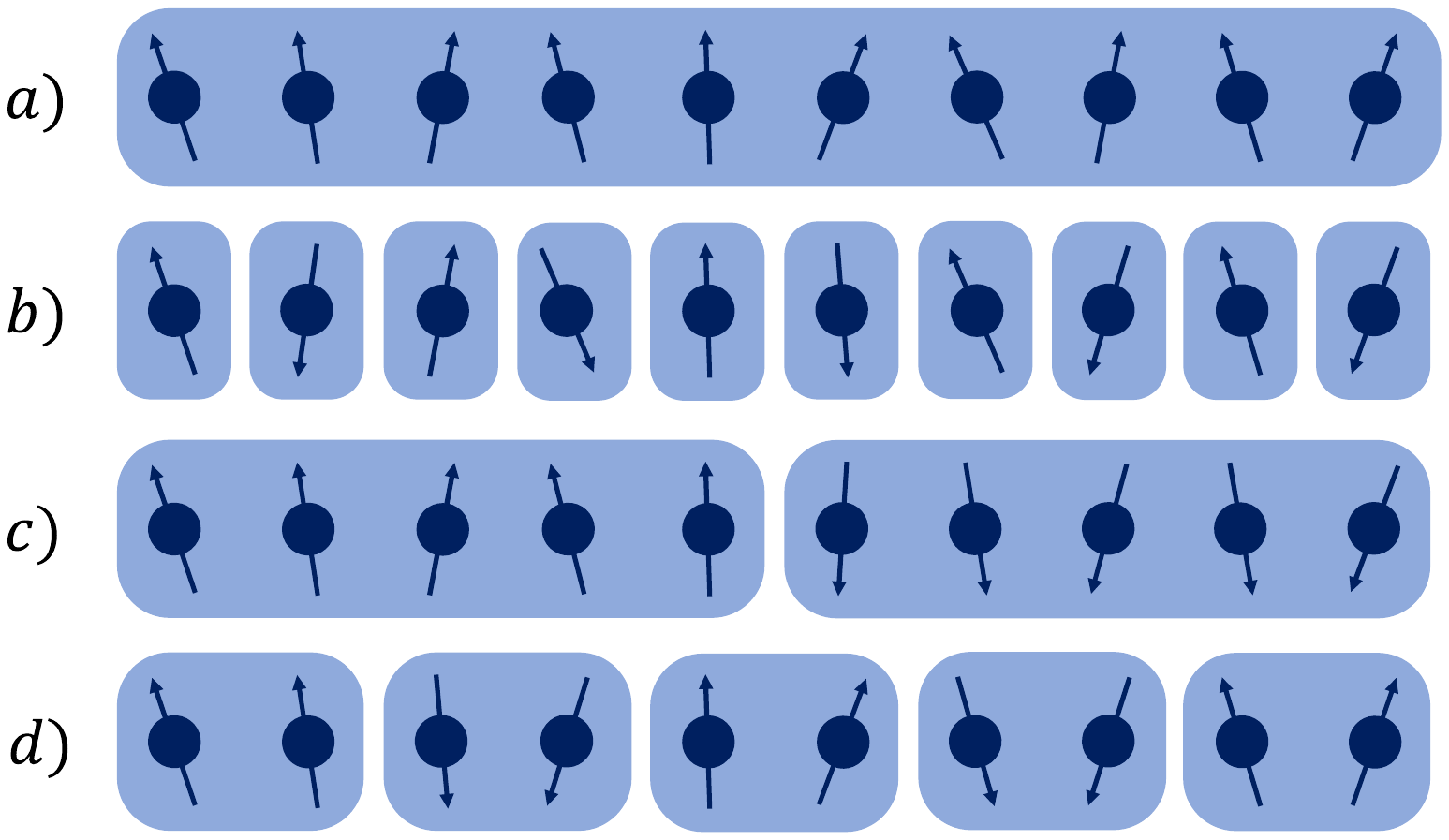}
\vspace{-13mm}
\caption{
Models for spin-$1/2$ chains coupled to independent reservoirs, whose range of interaction is depicted by the light blue shades over the spins.
a) Fully-global model: The entire chain is coupled to the same bath.
b) Fully-local case: Individual spins are coupled to their own local bath.
c-d) Intermediate bath-locality models: Each bath couples to more than a single spin with, e.g.,
 c) Half-and-half coupling and d) Pairwise coupling.
}
\label{fig:Fig1}
\end{figure}

%
After discussing the mapping approach,
we apply it on several spin models coupled globally or locally to different non-Markovian baths (super-Ohmic, Brownian)
and examine their equilibrium phases as a function of system-bath couplings at low temperature. 


\textit{Mapping.---}
We consider a many-body system described by the Hamiltonian $\hat{H}_S$ coupled to a 
bosonic-harmonic environment. 
For simplicity, we describe the mapping in a model with a single heat bath. 
The total Hamiltonian of the system, environment, and their interaction reads
\begin{equation}
   \begin{aligned} \hat{H}=&\hat{H}_S+\hat{H}_B+\hat{H}_I\\
=&\hat{H}_S+\sum_k\nu_k\hat{c}^\dagger_k\hat{c}_k+\hat{S}\sum_kt_k\left(\hat{c}^\dagger_k+\hat{c}_k\right),
   \end{aligned}
   \label{eq:H}
\end{equation}
where $\hat{c}^\dagger_k~(\hat{c}_k)$ are bosonic creation~(annihilation) operators with frequency $\nu_k$ for the $k$-th harmonic mode. 
$\hat{S}$ is an operator defined over the system's degrees of freedom, which couples to the reservoir with a coupling strength captured by the bath spectral density function, $K(\omega)=\sum_kt^2_k\delta(\omega-\nu_k)$. 

An ``effective"  Hamiltonian (EFFH) can be constructed by sequentially 
applying the reaction coordinate and polaron transformations onto the total Hamiltonian Eq.~\eqref{eq:H},
followed by a controlled truncation \cite{Nick_PRX,Nick_PRB,SM}. This mapping
is defined such that the coupling of the system to the reservoir
is made weaker in the new picture, while the effects of strong system-bath couplings are absorbed into the effective system's Hamiltonian \cite{Nazir18}. 
Post mapping, the effective (eff) Hamiltonian reads 
$\hat{H}^\text{eff}=\hat{H}^\text{eff}_S (\lambda, \Omega)+\hat{H}^\text{eff}_B+\hat{H}^\text{eff}_I$, and
we highlight the dependence of the effective system's Hamiltonian on $\lambda$ and $\Omega$. These parameters 
are functions of the original spectral function of the bath, $K(\omega)$ \cite{SM}. They can be interpreted as a system-bath interaction energy scale ($\lambda$) and a characteristic frequency ($\Omega$) of the bath, both corresponding to the original bath described in Eq.~\eqref{eq:H}. 
The effective system couples to a modified (residual) bath $\hat{H}^\text{eff}_B=\sum_k\omega_k\hat{b}^\dagger_k\hat{b}_k$ 
through $\hat{H}^\text{eff}_I=-\frac{2\lambda}{\Omega}\hat{S}\sum_kf_k(\hat{b}^\dagger_k+\hat{b}_k)$;
 $\hat{b}^\dagger_k~(\hat{b}_k)$ corresponds to new bosonic creation~(annihilation) operators with frequency $\omega_k$.
Importantly, a $\kappa$ scaling of the original coupling, $\kappa K(\omega)$, does not impact the spectral function of the residual bath, $K^\text{RC}(\omega)=\sum_kf^2_k\delta(\omega-\omega_k)$ \cite{Nazir18,Nick_PRB,Nick_PRX,SM}; 
the spectral function $K^{\text{eff}}(\omega)=\frac{4\lambda^2}{\Omega^2} K^{\text {RC}}(\omega)$
characterizes the effective bath, and we work in the limit of $\lambda\lesssim\Omega$.
%
%
%
%
%
The scaling observation allows building effective models in which the residual bath only weakly couples to the system~\cite{SM}.
This allows us to compute the system's equilibrium state resulting from its interaction with the bath as the Gibbs state with respect to the {\it effective} system's Hamiltonian, $\rho^\text{eff}_S=
\frac{1}{Z^\text{eff}}e^{-\beta\hat{H}^\text{eff}_S}$; 
$Z^\text{eff}=\Tr[e^{-\beta\hat{H}^\text{eff}_S}]$ is the partition function with $\beta$ the inverse temperature of the bath~\cite{Breuer,Cresser,Keeling2022}. This approach, was successfully validated on impurity models~\cite{Nick_PRB,Nick_PRX}, and it is utilized here as a general analytical method for tailoring magnetic order in open quantum lattices.

%

%
%

\textit{Spin chains.---}
We enact the mapping procedure on the general Heisenberg spin chain and its variants to expose bath-induced phase order.
The dissipative Heisenberg chain with $N$ sites is given by Eq.~\eqref{eq:H}, with the system's Hamiltonian 
\begin{equation}
\label{eq:h_xyz}
    \hat{H}_S = \sum_{i=1}^{N} \Delta_i \hat{\sigma}^z_{i} + \sum_{\alpha}\sum^{N-1}_{i=1}J_\alpha\hat{\sigma}^\alpha_i\hat{\sigma}^\alpha_{i+1}.
\end{equation}
Here, $\Delta_i>0$ represents the spin splitting of the $i$th spin. 
We set $J_\alpha>0$ as the uniform interaction strength between neighboring spins in the $\alpha=\{x,y,z\}$ direction. The chain is coupled to multiple bosonic baths, and we consider four scenarios, depicted in Fig.~\ref{fig:Fig1}: 
(a) \textit{fully-global} and
(b) \textit{fully-local} baths,
as well as
(c) \textit{half-and-half}
and (d) \textit{pairwise} 
coupling schemes.
Cases (a) and (d) are presented here; the discussion of the other two models can be found in Ref.~\cite{SM}.

We implement two complementary mapping procedures on spin chains, detailed in Ref.~\cite{SM}: 
(i) We build on the reaction coordinate mapping to adjust the system-bath boundary, followed by a polaron rotation of the reaction coordinate and its truncation \cite{Nick_PRX,Nick_PRB}. 
(ii) We apply a polaron rotation {\it directly} on the interaction Hamiltonian, acting on all modes in the bath. 
We show in Ref. \cite{SM} that the two mapping methods build completely analogous system's Hamiltonian
$\hat{H}^\text{eff}_S$, along with a weakened system-bath coupling strength. 
For equilibrium properties, the two approaches thus provide parallel results \cite{Nick_PRB};
deviations may show in time-dependent simulations. 
Conceptually, the methods can each handle general spectral density functions, yet
it is convenient to enact the first (i) mapping method on a Brownian bath
with $ K(\omega) = \frac{4\gamma \Omega^2\lambda^2\omega}{(\omega^2-\Omega^2)^2+(2\pi\gamma\Omega\omega)^2}$; $\lambda$ is the system-bath coupling energy and the bath is peaked at $\Omega$ with width energy $\gamma\Omega$.
In the effective picture, $K^{\text {eff}}(\omega)  \propto \gamma\omega$ \cite{Nick2021},
thus the system-bath interaction in $H^{\text {eff}}$ becomes weak once $\gamma \ll1$.
The second mapping (ii) can be readily performed on the Ohmic family spectral functions, e.g.,
$K(\omega)=\alpha\frac{\omega^3}{\omega^2_c}e^{-\omega/\omega_c}$ with $\alpha$ a dimensionless coupling parameter; under the polaron picture the parameters $\lambda$ and $\Omega$ that are used to define the effective system Hamiltonian can be expressed in terms of $\alpha$ and $\omega_c$ \cite{SM}.
\begin{figure}[htpb]
\fontsize{6}{10}\selectfont
\includegraphics[width=1\columnwidth]{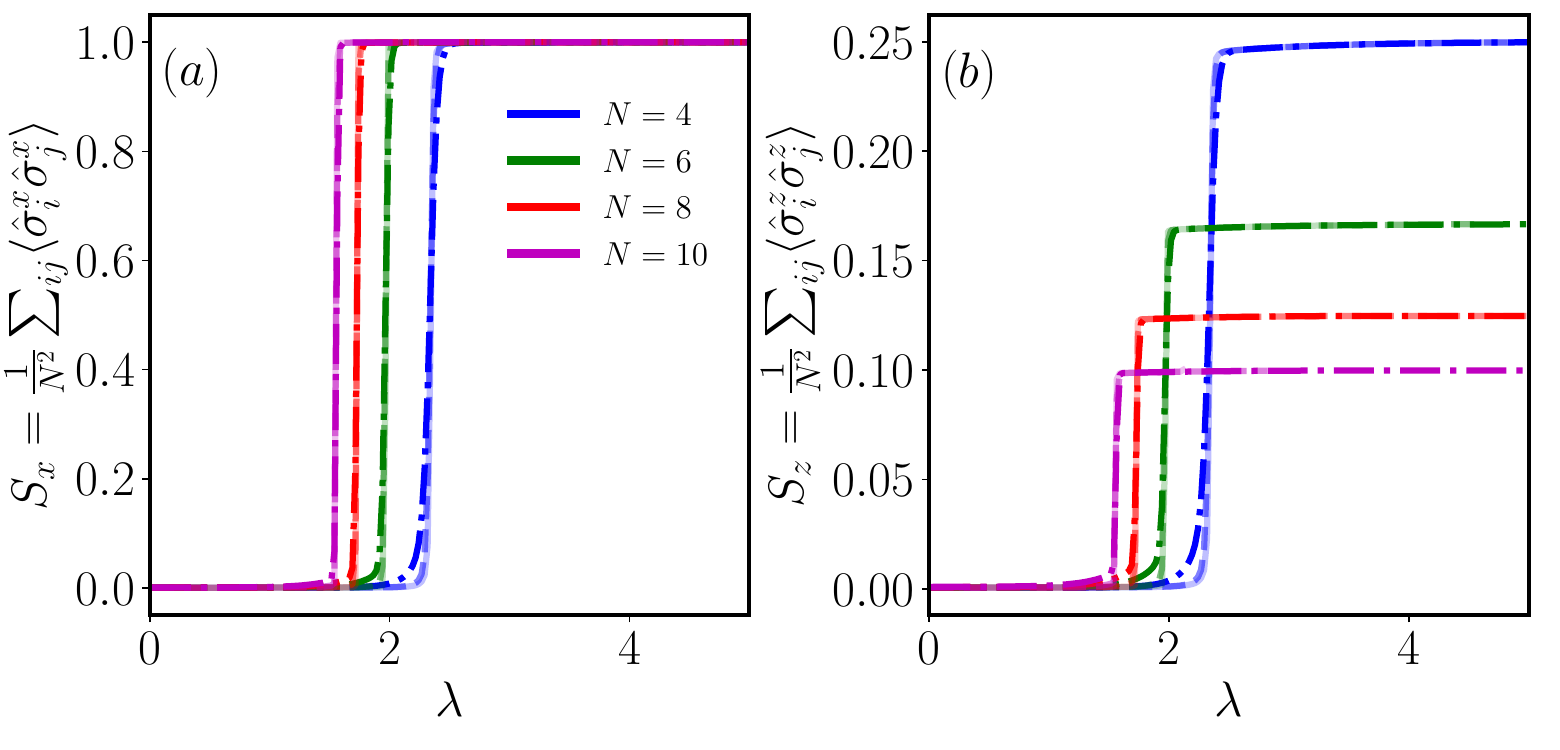}
\caption{Heisenberg spin chain in a global bath. We display the
structure factors $S_\alpha=\frac{1}{N^2}\sum_{ij}\langle \hat{\sigma}^\alpha_i\hat{\sigma}^\alpha_j\rangle$
in the (a) $\alpha=x$ and (b) $\alpha=z$ directions
as a function of the system-bath interaction energy, $\lambda$.
%
Other parameters are $\Delta=0.1$, $\Omega=10$, $J_x=1$, $J_y=0.9$, $J_z=0.8$.
We study chains with $N=\{4,6,8,10\}$ spins;  dashed-dotted, dashed, and solid lines (about overlapping) 
correspond to $T=0.2,0.1$, and $0.05$,  respectively. 
}
\label{fig:Sxyz}
\end{figure}

\begin{figure*}[htpb]
\fontsize{6}{10}\selectfont
\centering
\includegraphics[width=2\columnwidth]{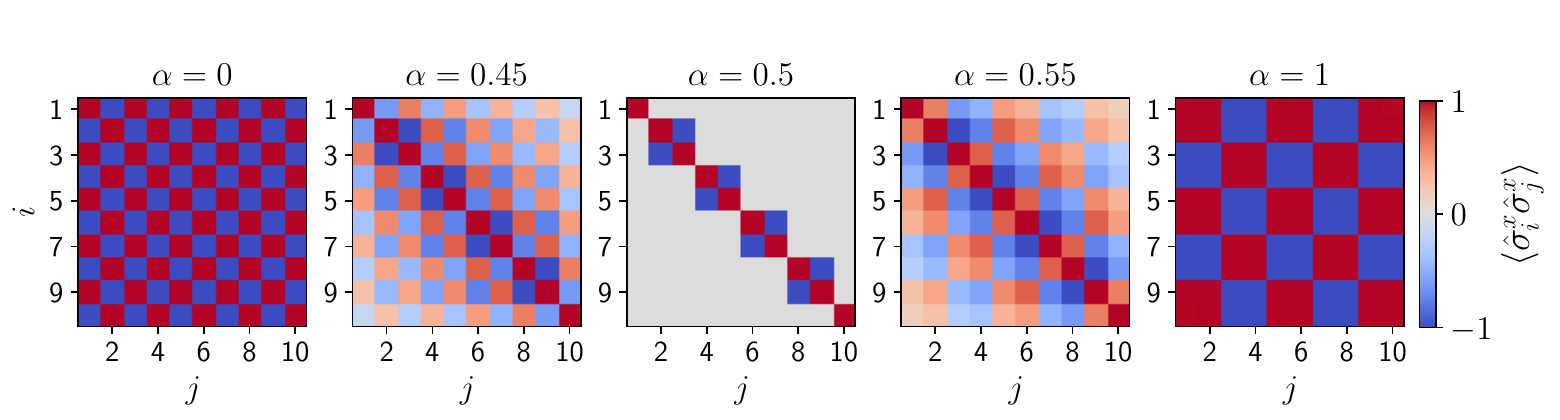}
\caption{Ising chain with pairwise couplings to baths.
We display spin-spin correlations $\langle \hat{\sigma}^x_i\hat{\sigma}^x_j\rangle$ for $N=10$, $\Delta=0.1$, $J_x=1$, $J_y=J_z=0$, 
$T=0.1$ and $\omega_c=0.5$. 
(left-to-right) We increase $\alpha$,
the dimensionless coupling parameter in the super-Ohmic bath model.
Values of $\alpha$ are chosen to manifest the crossover from an AFM order to an extended AFM order. Spin-spin correlations for spins coupled to the same bath precisely diminish at $\alpha=0.5$
where we lose all long-range correlations, with the two edge spins decoupled from the rest of the chain.
}
\label{fig:Fig3}
\end{figure*}


\textit{Fully-global coupling.---}
The Hamiltonian is given by  Eq.~\eqref{eq:H} with the Heisenberg Hamiltonian Eq.~\eqref{eq:h_xyz}. In the fully-global coupling model,
all the spins couple to a single bath and we use as an example the interaction operator 
$\hat{S}_\text{glob}=\sum^N_{i=1}\hat{\sigma}^x_i$.  
The mapped system is given by \cite{SM} 
\begin{equation}
\label{eq:HGeff}
\hat{H}^\text{eff}_{\text{glob},S} =\sum^N_{i=1}\tilde{\Delta}_i\hat{\sigma}^z_i+\sum_{\alpha}\sum^{N-1}_{i=1}\tilde{J}_\alpha\hat{\sigma}^\alpha_i\hat{\sigma}^\alpha_{i+1}-\frac{\lambda^2}{\Omega}\hat{S}^2_\text{glob}
\end{equation}
%
with $\tilde{\Delta}_i=\Delta_ie^{-\frac{2\lambda^2} {\Omega^2}}$, $\tilde{J}_x=J_x$, and 
spin interactions renormalized according to
\begin{equation}
    \tilde{J}_{y(z)} = \frac{J_{y(z)}}{2}\left(1+e^{-\frac{8\lambda^2}{\Omega^2}}\right)+\frac{J_{z(y)}}{2}\left(1-e^{-\frac{8\lambda^2}{\Omega^2}}\right).
\end{equation}
%
%
%
Let us examine some key features of the effective system Hamiltonian, Eq.~\eqref{eq:HGeff}. 
First, as expected, $\hat{H}^\text{eff}_{\text{glob},S}\rightarrow \hat{H}_S$ as $\lambda\rightarrow 0$. 
Second, environmental effects on the magnetic order at low temperature 
are transparent in this picture: As a function of $\lambda$, with our particular choice of $\hat{S}_\text{glob}$ 
(which the mapping is not limited to), 
the effect of the environment is to mix the anisotropies with respect the $x$ component, 
i.e., the $y$ and $z$ components. 
Furthermore, the individual spin splittings $\Delta_i$ are exponentially suppressed 
as $\lambda$ increases. This suppression can be rationalized 
as the entire chain is coupled in the $x$ direction, which leads to spins aligning 
in that direction as the coupling strength increases. 
One can imagine an analogous scenario of turning on a strong magnetic field in the $x$ direction, which
would similarly suppress spin components in the $z$ direction. 
Most dramatically, the last term in Eq.~\eqref{eq:HGeff} describes all-to-all spin 
interactions arising in the $x$ direction at nonzero $\lambda$. For later use, 
we denote this energy by $E_I\equiv \frac{\lambda^2}{\Omega}$.
In the super-Ohmic model, this bath-induced coupling is given by $E_I\equiv 2 \alpha \omega_c$ \cite{SM}.
On physical grounds, this term with its accompanied minus sign is to be expected since the spin chain is coupled to a common environment \cite{FM}. 
%
Recall that $\lambda$ and $\Omega$ can be derived for distinct baths' spectral density functions, ensuring the versatility of the mapping.

\begin{figure}[h]
\fontsize{6}{10}\selectfont 
\includegraphics[width=1\columnwidth]{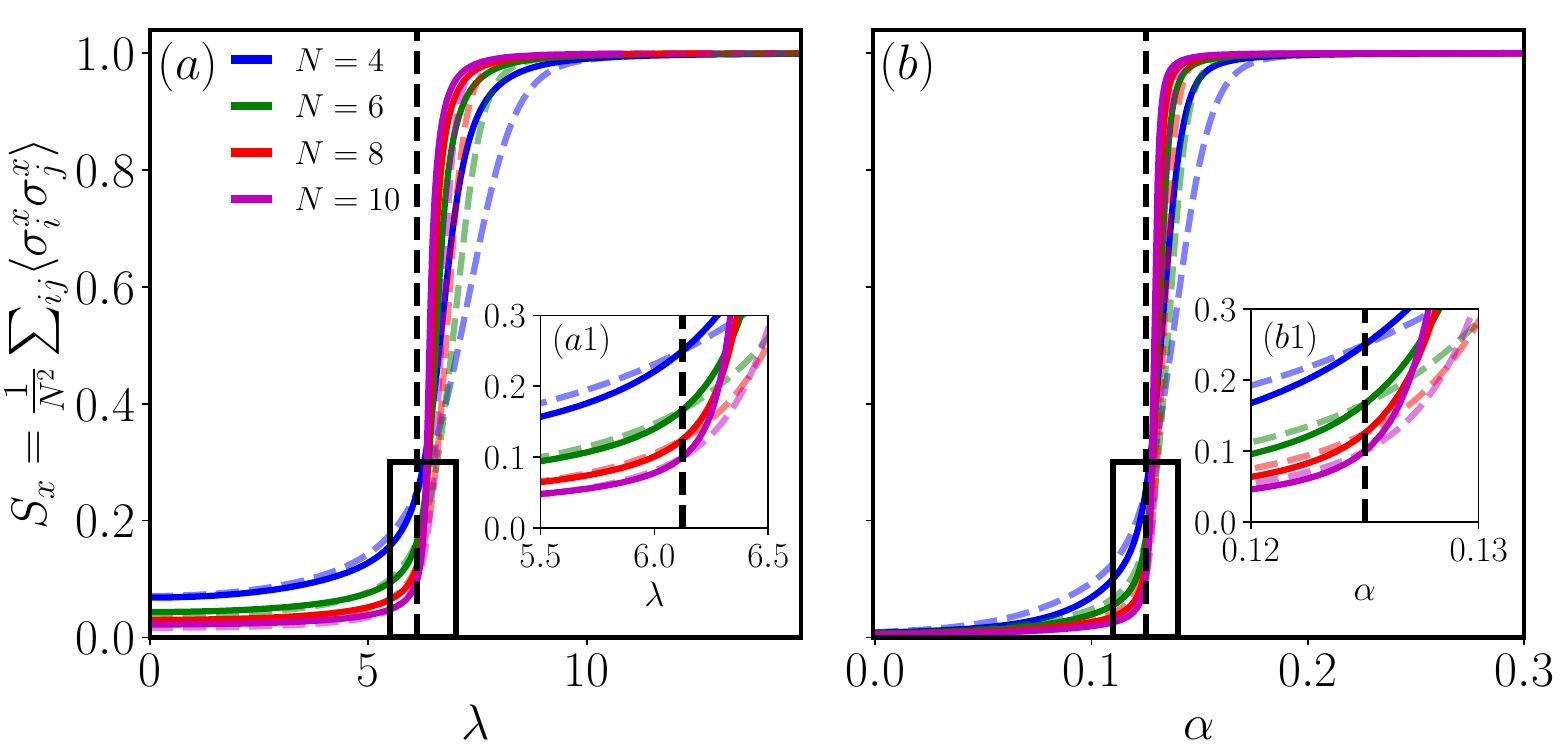}
\caption{Bath-induced quantum phase transition in the fully connected and globally coupled Ising model.
We present the structure factor 
$S_x=\frac{1}{N^2}\sum_{ij}\langle \hat{\sigma}^x_i\hat{\sigma}^x_j\rangle$ 
using parameters corresponding to (a) Brownian 
(b) and a super-Ohmic spectral functions. 
We  use $\Delta=0.1$, $J=3 (10)$ for $a(b)$, $\Omega=10$, and $\omega_c=0.5$.  
Insets (a1) and (b1) zoom over the corresponding main panels on the location of the quantum phase transition. 
Results are presented for $N=\{4,6,8,10\}$ 
at two temperatures $T=0.05$ (solid) and $0.1$ (dashed). 
The dashed black line indicates where the critical $(\lambda_c)~\alpha_c$ occurs,
 which corresponds to the point where the spin-spin interactions turn into ferromagnetic: 
Notably, the critical coupling strength is independent of temperature and chain length (insets).
}
\label{Fig:QPT}
\end{figure}

Corroborating these observations, deduced from Eq.~\eqref{eq:HGeff}, in Fig.~\ref{fig:Sxyz} we 
simulate the structure factor  $S_\alpha=\frac{1}{N^2}\sum_{ij}\langle\hat{\sigma}^\alpha_i\hat{\sigma}^\alpha_j \rangle$
for a Heisenberg chain with $N$ spins. 
The thermal average is done over the density matrix built from the effective Hamiltonian of the system, $\hat{H}^\text{eff}_{\text{glob},S}$.
The structure factor manifests a clear crossover with increasing $\lambda$, from the 
antiferromagnetic~(AFM) alignment of spins 
due to $J_\alpha>0$, to a FM order in the $x$ direction,
with $S_x$ going from $0$ to $1$ (Fig.~\ref{fig:Sxyz}(a)).
Furthermore, $S_z$ (and similarly $S_y$, not shown)
 demonstrate that all correlations in the $z$ (and $y$) directions are
 suppressed,  except  autocorrelators, thus reaching $1/N$ at strong coupling (Fig.~\ref{fig:Sxyz}(b)).
Few other comments are in place:
(i)  $S_\alpha=0$ at zero coupling due to the choice $J_\alpha>0$. 
(ii) To validate results, in \cite{SM} we benchmark the mapping technique against
the numerically accurate reaction-coordinate (RC) method \cite{NazirPRA14, Nazir16, Nazir18,Nick_2021,Nick2022,Felix2022,Nick2021,Nick_PRB}. We demonstrate an excellent agreement, particularly as $N$ grows, even at low temperature.
(iii) Given the collective nature of the coupling, the AFM to FM transition point will continue to shift to smaller $\lambda$ as $N$ grows. In contrast, the fully connected Ising model presented in Eq. (\ref{eq:HF}) supports a quantum phase transition at
a converged value of $\lambda>0$, independent of $N$, as we show in Fig. \ref{Fig:QPT}.



\textit{Pairwise coupling in the Ising chain.---} We examine next a simpler
version of the system Hamiltonian, Eq.~\eqref{eq:h_xyz}, by setting $J_y=J_z=0$, 
thereby making it a quantum Ising chain. We couple the chain to a collection of baths
as follows: every odd site of the chain, along with its nearest neighbour to the right, are coupled to a common bath as shown in Fig.~\ref{fig:Fig1}.~$d)$.
The total Hamiltonian is   
\bea
\hat{H}_\text{pair}  &=&
\hat{H}^\text{Ising}_S +\sum^{N/2}_{n=1} \hat{S}_\text{pair,n}\sum_kt_{n,k}\left(\hat{c}^\dagger_{n,k} 
+ \hat{c}_{n,k}\right)
\nonumber\\
&+&\sum_{n,k}\nu_{n,k}\hat{c}^\dagger_{n,k}\hat{c}_{n,k}.
\eea
Here, $\hat{H}^\text{Ising}_S=\hat{H}_S(J_y=J_z=0)$ 
and $\hat{S}_{\text{pair},n}=\hat{\sigma}^x_{2n-1}+\hat{\sigma}^x_{2n}$; $n\in\{1,\dots,N/2\}$ is the bath index. 
After the mapping \cite{SM}, the effective system Hamiltonian becomes
\bea
\label{eq:HP}
&&\hat{H}^\text{Ising,eff}_S=\sum^{N/2}_{n=1}\left(\bar{\Delta}_{2n-1}\hat{\sigma}^z_{2n-1}+\bar{\Delta}_{2n}\hat{\sigma}^z_{2n}\right)
\nonumber\\
  &&  +\sum^{N/2}_{n=1}\left(J_x-\frac{2\lambda^2_{2n-1}}{\Omega_{2n-1}}\right)\hat{\sigma}^x_{2n-1}\hat{\sigma}^x_{2n}+\sum^{N/2}_{n=1}J_x\hat{\sigma}^x_{2n}\hat{\sigma}^x_{2n+1} \,\,\,\,
\eea
where $\Bar{\Delta}_{2n-1}=\Delta_{2n-1}\exp(-2\lambda^2_n/\Omega^2_n)$ and $\bar{\Delta}_{2n}=\Delta_{2n}\exp(-2\lambda^2_n/\Omega^2_n)$.
We expect the two spins that are coupled to a common bath to build an FM alignment 
once the prefactor $\left(J_x-\frac{2\lambda^2_{2n-1}}{\Omega_{2n-1}}\right)$ becomes negative at sufficiently strong coupling $\lambda$.
In contrast, inter-cell interactions (between pairs)
continue to prefer an AFM alignment, captured by the last term in Eq.~\eqref{eq:HP}. 
The combination of these two effects creates an extended Neel order at sufficiently strong coupling, where at low temperature the preferred alignment is 
$\ket{\uparrow\uparrow\downarrow\downarrow\uparrow\uparrow\dots\uparrow\uparrow\downarrow\downarrow}$ 
in the $x$ direction, or the opposite case. 

In Fig.~\ref{fig:Fig3}, we display spin-spin correlations $\langle \hat{\sigma}^x_i\hat{\sigma}^x_j\rangle$ for an $N=10$-long chain. 
We clearly observe the buildup of spin alignments in subcells within the larger-scale AFM order as we increase the coupling parameter (left to right).
As an example, we assume here super-Ohmic spectral density functions for the baths (before the mapping).
As we show in \cite{SM}, the pairwise ferromagnetic coupling becomes then $E_I=2\alpha \omega_c$  (assuming identical baths).
Thus, with our choice of parameters ($J_x=1$, $\omega_c=5\Delta$), 
at $\alpha=0.5$, we precisely observe the complete suppression of long-range correlations at $J_x=2 \lambda^2_{2n-1}/\Omega^2_{2n-1}$.
Furthermore, at this value the two edge spins isolate from the rest of the chain---resulting from the segmentation of the chain into pairwise sectors.


\textit{Fully-connected Ising model.---} 
We now describe a model that exhibits a bath-induced quantum phase transition (QPT) at a particular coupling strength by allowing spins to interact beyond nearest neighbour in contrast to the Heisenberg model presented earlier. We return to model (a) in Fig.~\ref{fig:Fig1}, with a spin system globally-collectively coupled to a single common bath. The system's Hamiltonian assumed now is the fully-connected Ising model, reading
\begin{equation}
\label{eq:HF}
    \hat{H}_S = -\frac{\Delta}{2}\sum^N_{i=1}\hat{\sigma}^z_i+\frac{J\Delta}{8}\sum^N_{i,j=1}\hat{\sigma}^x_i\hat{\sigma}^x_j,
\end{equation}
where $\Delta>0$ is the spin splitting. Here, $J>0$ is a dimensionless parameter which scales the all-to-all spin interactions in the $x$ direction with respect to $\Delta$. 
This model exhibits a QPT of a Beretzinski-Kosterlitz-Thouless (BKT) type under Ohmic dissipation as demonstrated in Ref. \cite{De_Filippis_2021} via the quantum Monte Carlo technique.
 The mechanism behind this QPT is the bath induced FM interaction overcoming the intrinsic  AFM interaction $J$. 
Importantly, the critical coupling strength is system size {\it independent} once $J\neq0$,  which allows to identify the range of coupling strength that will retain the isolated-bath state 
even in the thermodynamic limit. 
This robustness contrasts the critical interaction scaling as $1/N$ when $J=0$. 

Our analytical mapping technique allows us to directly understand and predict this QPT from the effective Hamiltonian picture, and for general spectral functions. 
We couple the system \eqref{eq:HF} to a single bosonic bath and achieve the following effective system Hamiltonian \cite{SM}
\begin{equation}
    \hat{H}^\text{eff}_S = -\frac{\tilde{\Delta}}{2}\sum^N_{i=1}\hat{\sigma}^z_i+\left(\frac{J\Delta}{8}-\frac{\lambda^2}{\Omega}\right)\sum^{N}_{i,j=1}\hat{\sigma}^x_i\hat{\sigma}^x_j,
\end{equation}
Here, individual spin splittings $\Delta$ are suppressed to $\tilde{\Delta}$ in exactly the same manner as in Eq.~\eqref{eq:HGeff}.
However, unlike the Heisenberg chain with only nearest-neighbour interactions, 
 bath-induced FM interactions compete with the positive AFM interaction term $J\Delta$.  
Thus, while the example of Fig.~\ref{fig:Sxyz} 
displayed a monotonic shifting of the critical coupling strength to lower values as we increase $N$, 
in the fully-connected Ising model the critical coupling  converges to a constant value independent of $N$. 
In Fig.~\ref{Fig:QPT}, we demonstrate this critical bath coupling strength for the QPT by computing the structure factor $S_x$ for both Brownian $(a)$ and super-Ohmic $(b)$ baths.
The critical bath coupling $\lambda_c~(\alpha_c)$ is directly obtained at the points 
when the original AFM order shifts to a FM order: $\frac{J\Delta}{8}=\frac{\lambda^2_c}{\Omega}$
$(\frac{J\Delta}{8}=2\omega_c\alpha_c)$.  
Furthermore, we observe that the transition to a FM phase captured by $S_x$ 
becomes steeper with decreasing temperature as well as increasing number of spins, and we expect to see a discontinuous jump as $\beta,N\rightarrow\infty$ \cite{commentBKT}.

\textit{Discussion.---} 
We showed that an analytical mapping scheme yields clear insights on dissipative phase transitions in a broad class of spin systems, shedding light on 
phenomena that were previously approached independently, and with costly numerical tools.
The mapping takes a many-body spin Hamiltonian at potentially strong coupling to heat baths and transforms it into an effective spin model at {\it weak} coupling to (modified-residual) environments,  for which equilibrium expectation values can be readily evaluated using the Gibbs equilibrium state.
Specifically, we demonstrated that in Heisenberg chains
a global bath turns a low-temperature AFM order into a FM phase; 
an extended Neel phase is created when pairs of spins couple to a common bath; 
a bath-induced QPT occurs in the fully-connected Ising model. 
Regarding the validity of our results, one needs to operate in the regime where the system-bath coupling---in the effective Hamiltonian picture---remains weak. 
Furthermore, the reaction coordinate mapping assumes high-frequency baths \cite{Nick2021,RC}. 
As for temperatures, a comparison against more precise numerical tools \cite{SM} reveals that the mapping method progressively becomes {\it more accurate} with increasing chain length, even at low temperature. Additional studies are required to understand this encouraging trend.
The mapping approach was formulated for harmonic baths, but one can generalize it to other environments, including spin baths. The  scheme can also be readily generalized to higher spin systems and more complex system-bath operators. Moreover, the method lends itself to the analysis of bath-induced phases in disordered systems. 
With its generality and transparent form, the mapping method could be employed to design dissipation-controlled topological phases at finite temperature, the focus of our future work.


\begin{acknowledgments}
\textit{Acknowledgement.---}
We acknowledge fruitful discussions with Yuxuan Zhang, Yong-Baek Kim and Kartiek Agarwal. 
DS acknowledges support from an NSERC Discovery Grant and the Canada Research Chair program. 
The work of NAS was supported by Ontario Graduate Scholarship (OGS). 
The work of MB has been supported by the Centre for Quantum Information and Quantum Control (CQIQC) at the University of Toronto. 
Computations were performed on the Niagara supercomputer at the SciNet HPC Consortium. 
SciNet is funded by: the Canada Foundation for Innovation; 
the Government of Ontario; Ontario Research Fund - Research Excellence; and the University of Toronto.
\end{acknowledgments}

\bibliography{bibliography}

\end{document}


\title{Supplemental Material for ``Bath-engineering magnetic order in quantum spin chains: An analytic mapping approach"}

\author{Brett Min}
\email{brett.min@mail.utoronto.ca}
\affiliation{Department of Physics and Centre for Quantum Information and Quantum Control, University of Toronto, 60 Saint George St., Toronto, Ontario, M5S 1A7, Canada}

\author{Nicholas Anto-Sztrikacs}
\affiliation{Department of Physics and Centre for Quantum Information and Quantum Control, University of Toronto, 60 Saint George St., Toronto, Ontario, M5S 1A7, Canada}

\author{Marlon Brenes}
\affiliation{Department of Physics and Centre for Quantum Information and Quantum Control, University of Toronto, 60 Saint George St., Toronto, Ontario, M5S 1A7, Canada}

\author{Dvira Segal}
\email{dvira.segal@utoronto.ca}
\affiliation{Department of Chemistry,
University of Toronto, 80 Saint George St., Toronto, Ontario, M5S 3H6, Canada}

\affiliation{Department of Physics and Centre for Quantum Information and Quantum Control, University of Toronto, 60 Saint George St., Toronto, Ontario, M5S 1A7, Canada}

\date{\today}
\maketitle

\onecolumngrid

\renewcommand{\theequation}{S\arabic{equation}}
\renewcommand{\thefigure}{S\arabic{figure}}
\renewcommand{\thesection}{S\arabic{section}}
\setcounter{equation}{0}  
\setcounter{figure}{0}

This Supplemental Material includes the following: In Sec.~\ref{sec: Effective Hamiltonian Mapping}, we provide the details on the effective Hamiltonian mapping (EFFH) technique 
and its utility in studying the equilibrium state of a many-body system strongly coupled to its environment. In Sec.~\ref{sec: Application to the Dissipative Heisenberg chain}, we go through a detailed derivation of extracting the equilibrium state of a dissipative spin chain via this mapping technique. 
Importantly, we compare results of the EFFH methods to the more accurate numerical reaction coordinate technique, and validate the EFFH predictions. 
In Sec.~\ref{sec:Polaron}, we provide an alternative approach of obtaining the equilibrium state of the dissipative spin chain via the general polaron transform. 
We compare the two mapping approaches, the EFFH and the polaron, in Sec. \ref{sec:Disc}.
Finally, in Sec.~\ref{sec: Application to the Transverse field Ising chain}, we apply both 
the EFFH and the general polaron technique to the Ising chain to clearly visualize the preferred magnetic order in various coupling schemes depicted in the main text. 

\section{Effective Hamiltonian Mapping}
\label{sec: Effective Hamiltonian Mapping}

In this Section, we show how to transform a Hamiltonian into its ``effective" Hamiltonian model, such that in the new picture, the interaction of the system with the surrounding bath(s) can be made weaker than it was in the original model.
Remarkably, this transformation, which is not exact, 
immediately exposes the impact of system-bath couplings on the system in the form of renormalizing and mixing parameters, and in generating bath-mediated couplings.

The approach described here builds on Ref.~\cite{Nick_PRX,Nick_PRB}, where it was exercised on {\it impurity} models only such as the spin-boson model, a three level absorption refrigerator and a double quantum dot thermoelectric power generator. It is enacted here for the first time on spin chain models, and while varying the `range' of the bath(s), whether being global, local or of an intermediate locality range. 

The EFFH method builds on a reaction coordinate mapping, extracting a collective coordinate from the bath, followed by a polaron rotation of that mode and its truncation. In Sec. \ref{sec:Polaron} we discuss an alternative mapping approach that enacts polaron transform directly on all modes in the bath.

Consider a many-body system described by the Hamiltonian $\hat{H}_S$ coupled to a structured bosonic environment, modelled by an infinite set of harmonic modes. The total Hamiltonian of the system, environment and their interaction reads
\bea
\label{eq:total_h_vanilla}
\hat{H} &=& \hat{H}_S + \hat{H}_B + \hat{H}_I 
\nonumber\\
&=& \hat{H}_S + \sum_k \nu_k \hat{c}_k^{\dagger} \hat{c_k} + \hat{S}\sum_k t_k  \left( \hat{c}_k^{\dagger} + \hat{c}_k \right),
\eea
where the set $\{ \hat{c}_k \}$ are canonical bosonic operators, and $\nu_k$ and $t_k$ are the frequencies and the environment-system couplings between the $k$-th mode and the many-body system. $\hat{S}$ is an operator with support over system degrees of freedom that dictates the nature of the effect of the environment onto the system. Furthermore, the effect of the environment is typically described via spectral density function, which may be defined as $K(\omega) = \sum_k t^2_k \delta(\omega - \nu_k)$.

A reaction-coordinate (RC) mapping \cite{NazirPRA14,Nick2021} of Eq.~\eqref{eq:total_h_vanilla} leads to
\begin{align}
\label{eq:h_rc_tot}
\hat{H}_{\rm{RC}} = \hat{H}_S + \Omega \hat{a}^{\dagger} \hat{a} + \lambda \hat{S} \left( \hat{a}^{\dagger} + \hat{a} \right) + \sum_k f_k \left( \hat{a}^{\dagger} + \hat{a} \right) \left( \hat{b}_k^{\dagger} + \hat{b}_k \right) + \sum_k \omega_k \hat{b}_k^{\dagger} \hat{b}_k,
\end{align}
where $\{ \hat{a} \}$ is a canonical bosonic operator for the reaction coordinate with frequency $\Omega$. The parameter $\lambda$ now describes the coupling between the original system $\hat{H}_S$ and the reaction coordinate via the system operator $\hat{S}$. The enlarged system comprising $\hat{H}_S$ and the reaction coordinate are now coupled to a {\it residual} bath with the effect described by the new spectral function $K^{\rm RC}(\omega) = \sum_k f^2_k \delta(\omega - \omega_k)$. The harmonic modes of the residual reservoir are described via the canonical operators $\{ \hat{b}_k \}$ with frequencies $\omega_k$, which may be expressed as linear combinations of the original harmonic modes $\{ \hat{c}_k \}$. We note that both $\lambda$ and $\Omega$ follow from the original spectral density function, via $\lambda^2 = \frac{1}{\Omega} \int_0^\infty {\rm{d}}\omega \; \omega K(\omega)$ and $\Omega^2 = \frac{\int_0^\infty {\rm{d}}\omega \; \omega^3 K(\omega)}{\int_0^\infty {\rm{d}}\omega \; \omega K(\omega)}$ \cite{Nazir18}. 
%
Furthermore, the spectral density of the redefine-residual bath is $K^\text{RC}(\omega)=\frac{2\pi\lambda^2 K(\omega)}{[\mathcal{P}\int\frac{K(\omega')}{\omega'-\omega}d\omega']^2+\pi^2K^2(\omega)}$ with $\mathcal{P}$ indicating a principal-value integral \cite{Nazir18,NazirPRA14,Nick2021}. 

Next, we employ the polaron transformation on the RC mode, $\hat{\tilde{H}} = \hat{U} \hat{H}_{\rm{RC}} \hat{U}^{\dagger}$ with $\hat{U} = {\rm{exp}}\left[\frac{\lambda}{\Omega}(\hat{a}^{\dagger} - \hat{a})\hat{S}\right]$, such that
\begin{align}
\hat{\tilde{H}} = \hat{\tilde{H}}_S + \Omega \hat{a}^{\dagger} \hat{a} - \frac{\lambda^2}{\Omega}\hat{S}^2 + \sum_k f_k \left( \hat{a}^{\dagger} + \hat{a} - \frac{2\lambda}{\Omega} \hat{S} \right) \left( \hat{b}_k^{\dagger} + \hat{b}_k \right) + \sum_k \omega_k \hat{b}_k^{\dagger} \hat{b}_k.
\end{align}
Note that the polaron transform was enacted on a single mode, the reaction coordinate (which represents a collective mode of the original bath) and the system Hamiltonian. 

We  now perform an approximation by considering only the ground-state level of the reaction coordinate, by projecting $\hat{\tilde{H}}$ onto this manifold via the projector $\hat{Q}_0 = \ket{0}\bra{0}$, to obtain \cite{Nick_PRX}
\begin{align}
\label{eq:h_eff_total_projected}
\hat{H}^{\rm eff} &= \hat{Q}_0 \hat{\tilde{H}}_S \hat{Q}_0 - \frac{\lambda^2}{\Omega}\hat{S}^2 - \hat{S}\sum_k \frac{2\lambda f_k}{\Omega}  \left( \hat{b}_k^{\dagger} + \hat{b}_k \right) + \sum_k \omega_k \hat{b}_k^{\dagger} \hat{b}_k.
\end{align}
Crucially, Eq.~\eqref{eq:h_eff_total_projected} has the same form as the original total Hamiltonian $\hat{H}$ from Eq.~\eqref{eq:total_h_vanilla} with the new effective system Hamiltonian defined as 
\begin{align}
\label{eq:h_eff_sys}
\hat{H}^{\rm eff}_S = \hat{Q}_0 \hat{\tilde{H}}_S \hat{Q}_0 - \frac{\lambda^2}{\Omega}\hat{S}^2.
\end{align}
We also provide a useful identity that allows a quick extraction of the effective system Hamiltonian, generalizing Ref. \cite{Nick_PRX}:
\begin{equation}
\label{eq: effective dictionary}
    \hat{H}^\text{eff}_S = e^{-(\lambda^2/2\Omega^2)\hat{S}^2}\left(\sum^\infty_{n=0}\frac{\lambda^{2n}}{\Omega^{2n}n!}\hat{S}^n\hat{H}_S\hat{S}^n\right)e^{-(\lambda^2/2\Omega^2)\hat{S}^2} - \frac{\lambda^2}{\Omega}\hat{S}^2.
\end{equation}
Let us now discuss some key aspects of the transformed Hamiltonian described in Eq.~\eqref{eq:h_eff_total_projected}. 
First, note the coupling of the system to the modes of the residual bath is re-scaled $f_k \to 2\lambda f_k / \Omega$. If this residual coupling can be made a perturbative parameter, strong coupling effects can be studied as a function of $\lambda$. This constitutes a specific form of a Markovian embedding, where strong-coupling effects become embedded onto the effective system Hamiltonian. 
Furthermore, the spectral density of the effective bath will be further dressed by the RC parameters: $K^\text{eff}(\omega)=\frac{4\lambda^2}{\Omega^2}K^\text{RC}(\omega)$.
Worthy of note, for the case of an open system initially coupled to a bosonic bath with Brownian form, 
\bea K(\omega) = \frac{4\gamma\Omega^2 \lambda^2\omega}{(\omega^2-\Omega^2)^2 + (2\pi\gamma\Omega\omega)^2},
\label{eq:Brown}
\eea 
the effective spectral density function maps to $K^{\rm{eff}}(\omega) = \frac{4\lambda^2}{\Omega^2} \gamma \omega$, where $\gamma$ initially played the role of a width parameter, now represents the residual coupling strength between the residual bath and open system. 
However, the mapping is not limited to the Brownian function.
In fact, one can treat Eq.~\eqref{eq: effective dictionary}
as a general effective Hamiltonian emerging from Eq.~\eqref{eq:total_h_vanilla}
with $\Omega$ as the  frequency of the bath and $\lambda$ as a measure for the system-bath coupling energy. 
From Eq.~\eqref{eq: effective dictionary} we note that strong coupling effects (large $\lambda$) are now embedded in the effective system Hamiltonian, while the coupling to the residual bath can be made weak, by assuming $\gamma\ll1$ in the above mentioned Brownian example.
We also note that the coupling operator $\hat{S}$ has intricate consequences in relation to environmental effects. Indeed, while thermodynamic equilibration between a system and environment is a phenomenon that occurs irrespective of the microscopic details of the nature of the coupling between each subsystem; at strong coupling, on the other hand, it has been established that microscopic details are of the essence \cite{Cresser,Keeling2022}.

The effective Hamiltonian treatment allows one to understand equilibrium thermodynamics in many-body systems coupled to thermal environments. Through the Markovian embedding of the reaction-coordinate mapping, we conjecture 
that the equilibrium state of the many-body system is a Gibbs state of the effective Hamiltonian 
\bea
\hat{\rho}^{\rm eff}_S = e^{-\beta \hat{H}^{\rm eff}_S} / Z^{\rm eff}.
\label{eq:MFGS}
\eea
Here, $Z^{\rm eff} = \text{Tr}\left[e^{-\beta \hat{H}^{\rm eff}_S}\right]$ is the partition function of the effective system. It should be noted that at sufficiently strong $\lambda$, this state does not coincide with the thermal state of the isolated system, $\hat{H}_S$, and strong-coupling effects may be observed in the many-body equilibrium state. Indeed, $\hat{H}^{\rm eff}_S$ already contains terms that depend on $\lambda$ that generate interactions among all the spins composing the many-body system.
The equilibrium state in Eq.~\eqref{eq:MFGS} was analyzed in Refs.~\cite{Nick_PRB,Nick_PRX} for the spin-boson model, where we demonstrated that it provided an excellent approximation for equilibrium expectation values from the weak to the strong coupling limit. Moreover, it can be shown that the equilibrium state of the EFFH method is an {\it exact} result for the spin boson model in the ultrastrong coupling regime \cite{Nick_PRX}. 

\section{Application of the EFFH mapping to the Dissipative Heisenberg chain}
\label{sec: Application to the Dissipative Heisenberg chain}
In this Section, we apply the effective Hamiltonian mapping as described in Sec.~\ref{sec: Effective Hamiltonian Mapping} to the general dissipative spin-chain $\hat{H}_S$ given by the following Hamiltonian:
\begin{equation}
\label{eq: Heisenberg chain}
    \hat{H}_S = \sum_{i=1}^{N} \Delta_i \hat{\sigma}^z_{i} + \sum_{\alpha}\sum^{N-1}_{i=1}J_\alpha\hat{\sigma}^\alpha_i\hat{\sigma}^\alpha_{i+1},
\end{equation}
where $\Delta_i>0$ represents the spin splitting of the $i$th qubit, $J_\alpha>0$ is the interaction strength between neighbouring spins in the $\alpha=\{x,y,z\}$ direction. In the following subsections, we will examine the structure of the equilibrium state under four different dissipative coupling schemes which corresponds to \textit{Fully-global}~\ref{subsec: Fully-global eff}, \textit{Fully-local} \ref{subsec: Fully-local eff}, \textit{Half-and-half} \ref{subsec: Half-and-half eff}, and \textit{Pairwise} \ref{subsec: Pairwise eff} coupling schemes. 

As for notation, we denote the original spin sites by $i$ and $j$, and index baths with $n$. 
However, once the mapping is operated and depending on the type of the bath, we use notation
as convenient and clarify it in the text.

In the global model only one bath couples to the system. In the fully local model, the number of independent baths equals the number of sites. In the half-and-half model, the chain couples to two independent baths, while in the pairwise model
there are $N/2$ independent baths, with $N$ spins in the chain. 
These models are depicted in Fig.~1 in the Main text.

\subsection{Fully-global coupling model and Benchmarking}
\label{subsec: Fully-global eff}

We imagine the entire spin chain to be coupled to a single bosonic bath described by the following Hamiltonian,
\begin{equation}
    \hat{H}_\text{glob} = \hat{H}_S+\hat{S}_\text{glob}\sum_kt_k\left(\hat{c}^\dagger_k+\hat{c}_k\right)+\sum_k\nu_k\hat{c}^\dagger_k\hat{c}_k.
\end{equation}
Here, $\hat{H}_S$ is given by Eq.~(\ref{eq: Heisenberg chain}) and $\hat{S}_\text{glob}=\sum^N_{i=1}\hat{\sigma}^x_i$; $i$ is an index of spin sites in the chain.
We apply the EFFH framework to the global spin-chain following the framework outlines in Sec.~\ref{sec: Effective Hamiltonian Mapping}, extracting a single reaction coordinate from the global bath, polaron transforming it, then truncting it. 
A crucial point in the derivation is that $\hat{S}_\text{glob}^2 = N\hat{I} + 2\sum_{i<j} \hat{\sigma}^x_i\hat{\sigma}^x_j$, which implies that we generate new interactions between spins in the system due to the influence of the global reservoir. Following through the steps outlined in the previous section~\ref{sec: Effective Hamiltonian Mapping}, we arrive at the following effective system Hamiltonian, 
\bea
\label{eq:h_eff_tilted_heisenberg}
\hat{H}^{\rm eff}_{\text{glob},S} = \sum_{i=1}^N \tilde{\Delta}_i\hat{\sigma}^i_z + \sum_{i=1}^{N-1} (J_x \hat{\sigma}^x_{i} \hat{\sigma}_{i+1}^{x} + \tilde{J}_{y} \hat{\sigma}_i^{y} \hat{\sigma}_{i+1}^{y} + \tilde{J}_{z} \hat{\sigma}^z_{i} \hat{\sigma}^z_{i+1}) - \frac{\lambda^2}{\Omega}N \hat{I} - \frac{2\lambda^2}{\Omega}\sum_{i<j} \hat{\sigma}_i^x\hat{\sigma}_j^x,
\eea
where, $\tilde{\Delta}_i = \Delta_i e^{-\frac{2\lambda^2}{\Omega^2}}$ and 
\bea
\tilde{J}_{y} &= \frac{J_y}{2} \left( 1 + e^{-8\lambda^2/\Omega^2} \right) + \frac{J_z}{2} \left( 1 - e^{-8\lambda^2/\Omega^2} \right), \nonumber \\
\tilde{J}_{z} &= \frac{J_z}{2} \left( 1 + e^{-8\lambda^2/\Omega^2} \right) + \frac{J_y}{2} \left( 1 - e^{-8\lambda^2/\Omega^2} \right).
\eea
After the mapping, the effective system weakly couples to the residual bath, thus we can calculate equilibrium expectation values using the equilibrium state in Eq.~\eqref{eq:MFGS}. We do not write here explicitly the residual bath's Hamiltonian and the EFFH system-bath interaction term, which are included in Eq.~\eqref{eq:h_eff_total_projected}, since they do not play a role in the equilibrium state of the effective model.

Before showing any calculations, already at the level of the Hamiltonian described in Eq.~\eqref{eq:h_eff_tilted_heisenberg}, we can see that the spins should manifest a crossover in the $x$ direction with
increasing $\lambda$, from the anti-ferromagnetic (AFM) alignment of spins due to $J_x > 0$, to a FM order. 
This transition takes place due to the last negative term in the effective system's Hamiltonian, Eq.~\eqref{eq:h_eff_tilted_heisenberg}. Importantly, since this long-range term scales as $N^2$, with $N$ the total number of spins, the value of $\lambda$ at which the AFM to FM transition occurs will keep shifting to lower values with increasing the size $N$. This is in contrast to the fully-connected chain (with $J_x$ connecting all spins) where a fixed critical point takes place, at a specific $\lambda$. The fully-connected model is described in the main text. 

\begin{figure}[t]
\fontsize{13}{10}\selectfont 
\centering
\includegraphics[width=1\columnwidth]{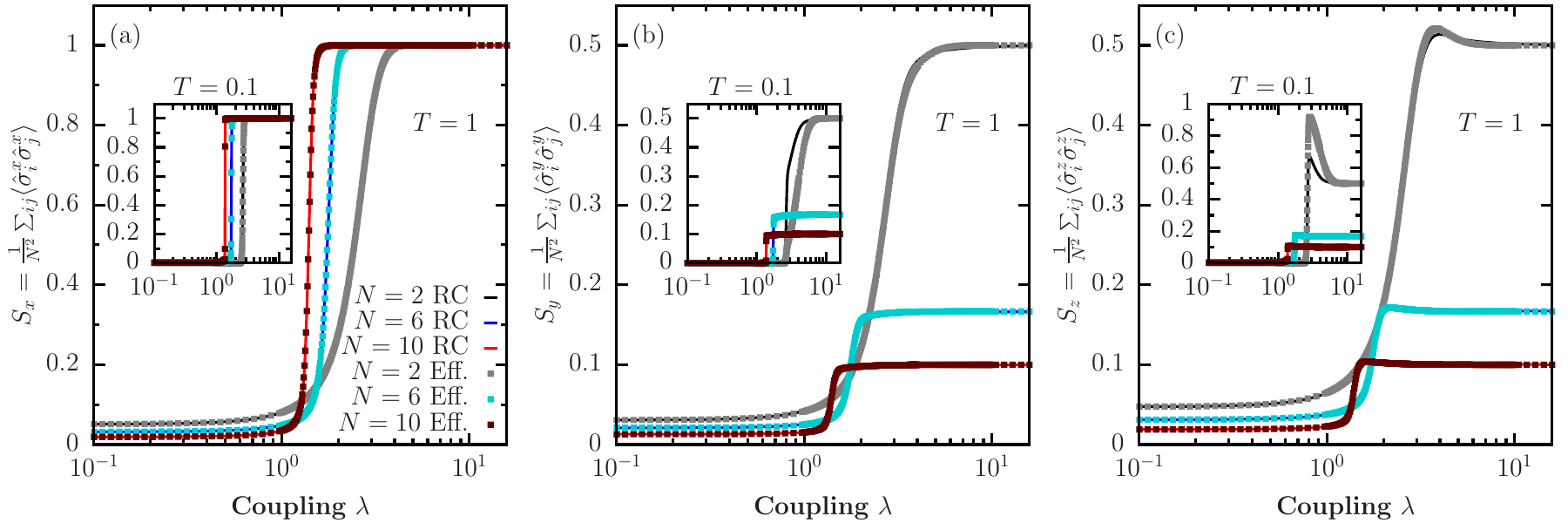}
\caption{Comparison of the expectation values of the spin structure factor $S_{\alpha} = \frac{1}{N^2} \sum_{ij} \langle \hat{\sigma}^{\alpha}_i \hat{\sigma}^{\alpha}_j \rangle$ for $\alpha=\{x,y,z\}$, computed from both the numerical $\hat{\rho}^{\rm RC}_S$ and the analytical $\hat{\rho}^{\rm eff}_S$ for different system sizes $N$ of the anisotropic Heisenberg chain. The parameters of the calculation are $J_x = 0.77$, $J_y = 1.23$, $J_z = 0.89$ and $\Omega = 8$. The main panels display the results for $T = 1$, while the insets show results for $T = 0.1$. With the increasing coupling strength $\lambda$, we find a crossover from $S_\alpha=0$ to $S_\alpha\neq0$, that becomes more steeper at lower temperature. The limiting value for $S_x$ becomes $1$ as the entire chain prefers a ferromagnetic alignment in the $x$ direction while the limiting value for both $S_y$ and $S_z$ tends to $1/N$ as the entire spin-spin correlations in both $y$ and $z$ directions are completely suppressed except the autocorrelators. We also note that the EFFH results show better agreement with the numerically accurate RC method at large $N$.}
\label{fig:1}
\end{figure}

To validate these predictions, based on the EFFH equilibrium state approximation, in Figure~\ref{fig:1}, we show a comparison between the structure factor $S_\alpha=\frac{1}{N^2}\sum_{ij}\langle\hat{\sigma}^\alpha_i\hat{\sigma}^\alpha_j\rangle$ for $\alpha=x,y,z$ of a general Heisenberg $N=10$-site chain computed from the EFFH and the numerically accurate reaction coordinate (RC) method \cite{NazirPRA14,Nick2021}. We choose generic parameters,  $\Delta_i=0$, $J_x=0.77$, $J_y=1.23$, $J_z=0.89$, and $\Omega=8$. 
For RC simulations, we used a Brownian bath for the original Hamiltonian, which is converted to an Ohmic bath in the RC picture. 
Importantly, we observe in Figure~\ref{fig:1} an excellent agreement between the equilibrium state EFFH and the RC numerical method. We also find that the accuracy of the EFFH method becomes better as $N$ increases.  
Note that prediction of the RC numerical method were benchmarked against numerically exact Hierarchical Equation of Motion simulations in Refs.~\cite{Nazir16,Nick_PRB}. 
In the first row of Fig. \ref{fig:heatmap} we further present a heat map of this crossover using a super-Ohmic spectral function as an example, with details explained in Sec. \ref{sec: Application to the Transverse field Ising chain}. 

%
\subsection{Fully-local coupling model}
\label{subsec: Fully-local eff}
Next, we move on to a fully-local coupling scheme where each spin couples to their own independent bosonic bath. 
The total Hamiltonian of the model is
\begin{equation}
    \hat{H}_\text{local} = \hat{H}_S+\sum^N_{n=1}\hat{S}_\text{local,n}\sum_kt_{n,k}\left(\hat{c}^\dagger_{n,k}+\hat{c}_{n,k}\right)+\sum_{n,k}\nu_{n,k}\hat{c}^\dagger_{n,k}\hat{c}_{n,k}.
\end{equation}
%
Here, $\hat{H}_S$ is again given by Eq.~\eqref{eq: Heisenberg chain} while $\hat{S}_{\text{local},n}=\hat{\sigma}^x_n$. That is, unlike the fully-global coupling scheme studied in the previous subsection~\ref{subsec: Fully-global eff}, $\hat{S}_{\text{local},n}$ acts on the $n$th individual site. 
Here, the index $n$ that counts the baths is identical to the site index $i$. 
Upon proceeding with the EFFH mapping,
we extract a reaction coordinate from each independent bath, polaron transform each of these collective modes and truncate them. 
The results is that there are no bath-induced spin-spin correlations via the 
$\hat{S}^2_{\text{local},n}$ term. 
The effective system Hamiltonian instead transforms to
\bea
\hat{H}^{\rm{eff}}_{\text{local},S} = && \sum_{n=1}^N \tilde{\Delta}_{n} 
\hat{\sigma}^z_{n} 
+ \sum_{i=1}^{N-1} \left[J_{x} \hat{\sigma}^x_{n} \hat{\sigma}^x_{n+1} %
+ \tilde{J}_y^{(n,n+1)} 
\hat{\sigma}^y_{n} \hat{\sigma}^y_{n+1}  
+
\tilde{J}_z^{(n,n+1)} 
\hat{\sigma}^z_{n} \hat{\sigma}^z_{n+1} 
\right],
\label{eq:H2Qeff}
\eea
where now the transformed parameters map as $\tilde{\Delta}_{n} = \Delta_{n} e^{-\frac{2\lambda_{n}^2}{\Omega_{n}^2}}$,
$\tilde{J}_y^{(n,n+1)} = J_y e^{-\frac{2\lambda_{n}^2}{\Omega_{n}^2}} e^{-\frac{2\lambda_{n+1}^2}{\Omega_{n+1}^2}}$, and
$\tilde{J}_z^{(n,n+1)} = J_z e^{-\frac{2\lambda_{n}^2}{\Omega_{n}^2}} e^{-\frac{2\lambda_{n+1}^2}{\Omega_{n+1}^2}}$, while $J_x$ stays intact.
We comment that the interactions with each reservoir can be made distinct, which offers flexibility in engineering the effective system, as the values of the parameters on each site can be tuned by the coupling. 

Furthermore, in contrast to the global scheme, there are only nearest-neighbour effects within this local bath picture. As the coupling to the baths increases, we expect the model to go from a Heisenberg chain to a quantum Ising model. This is due to the suppression of both of the $J_y$ and $J_z$ interaction terms with increasing coupling to the bath. 
In the second row of Fig. \ref{fig:heatmap} we present a heat map of this crossover, with details explained in Sec.~\ref{sec: Application to the Transverse field Ising chain}. 
\subsection{Half-and-half coupling model}
\label{subsec: Half-and-half eff}
We now demonstrate that by tuning the locality of the baths we can engineer different magnetic orderings in a spin chain. One simple example is a chain of $N$ sites, in which the left-half of the chain couples to one bath and the right-half of the chain couples to a separate bath. This model is described by the following Hamiltonian
\begin{equation}
    \hat{H}_\text{half} = \hat{H}_S+\sum_n\hat{S}_{\text{half},n}\sum_kt_{n,k}\left(\hat{c}^\dagger_{n,k}+\hat{c}_{n,k}\right)+\sum_{n,k}\nu_{n,k}\hat{c}^\dagger_{n,k}\hat{c}_{n,k},
\end{equation}
where $\hat{H}_S$ is again given by Eq.~\eqref{eq: Heisenberg chain}, while $n=\{L,R\}$, with $\hat{S}_{\text{half},L}=\sum^{N/2}_{i=1}\hat{\sigma}^x_i$, and $\hat{S}_{\text{half},R}=\sum^N_{N/2+1}\hat{\sigma}^x_i$. Crucially, now there are two different coupling depending on which half of the chain the spin is coupled to. As a result, the EFFH approach yields an intermediate scheme between the local and global approaches. For instance, in analogy with the global case, we see the generation of bath-induced long-ranged spin-spin interaction term via $\hat{S}_n^2 = \frac{N}{2}\hat{I} + 2\sum_{i<j} \hat{\sigma}^x_i\hat{\sigma}^x_j$ (where indices $i$ and $j$ should run appropriately depending on $n$, whether it is the left or right half). On top of it, we expect a suppression of both $J_y$ and $J_z$ spin-spin interaction for the boundary spins at the right~(left)-end of the left~(right) half of the chain. 

Technically, we extract a reaction coordinate mode from each of the two baths.
The polaron transformation involved in the mapping is made of two transformations that commute with each other,
\bea
\label{eq:polaron_transformation}
    \hat{U} = {\rm{exp}}\left[\frac{\lambda_L}{\Omega_L}(\hat{a}_L^{\dagger} - \hat{a}_L) \sum_{i=1}^{N/2} \hat{\sigma}^x_i  \right] {\rm{exp}}\left[\frac{\lambda_R}{\Omega_R}(\hat{a}_R^{\dagger} - \hat{a}_R) \sum_{i=N/2}^{N} \hat{\sigma}^x_i  \right] = \hat{U}_L\hat{U}_R,
\eea
where $\hat{a}^\dagger_n~(\hat{a}_n)$ corresponds to the creation~(annihilation) operator of the RC extracted from left ($n=L$) or right $(n=R)$ bath. We emphasize the necessity of $[\hat{U}_L,\hat{U}_R]=0$, which allows the EFFH method to proceed with minor modifications, similarly to the previous two examples. We note that the transformation for sites $i=1$ to $i=N/2 - 1$ and $i=N/2 + 2$ to $i=N$ will follow analogously to the global bath, while terms at boundary, sites $i=N/2$ and $i=N/2+1$ yield a transformation akin to the local bath. Using the results from the previous two sections, we find that the effective Hamiltonian of the system can be decomposed into a portion for the left bath (spin $1$ to $N/2$), the right bath (spin $N/2 + 1$ to $N$) and the boundary (spin $N/2$ and $N/2 + 1$) as 
\bea
    \hat{H}^{\rm eff}_{\text{half},S} = \hat{H}^{\rm eff}_{L,S} + \hat{H}^{\rm eff}_{\text{bound},S} + \hat{H}^{\rm eff}_{R,S}.
    \label{eq:HSH}
\eea
It follows that from the mapping, for each partition of the Hamiltonian we obtain 
\bea
    \hat{H}^{\rm eff}_{\text{half},L} = \sum_{i=1}^{N/2} \tilde{\Delta}_i\hat{\sigma}^z_i + \sum_{i=1}^{N/2-1} J_x \hat{\sigma}^x_{i} \hat{\sigma}^x_{i+1} + \tilde{J}_{y} \hat{\sigma}^y_{i} \hat{\sigma}^y_{i+1} + \tilde{J}_{z} \hat{\sigma}^z_{i} \hat{\sigma}^z_{i+1} - \frac{N}{2}\frac{\lambda_L^2}{\Omega_L} \hat{I} - \frac{2\lambda_L^2}{\Omega_L}\sum_{i<j = 1}^{N/2} \hat{\sigma}^x_i\hat{\sigma}^x_j. 
\eea
We note that the spin gets suppress as $\tilde{\Delta}_i = \Delta_i e^{-\frac{2\lambda_L^2}{\Omega_L^2}}$, and the interactions via $\tilde{J}_{y(z)} = \frac{J_{y(z)}}{2} \left( 1 + e^{-8\lambda_L^2/\Omega_L^2} \right) + \frac{J_{z(y)}}{2} \left( 1 - e^{-8\lambda_L^2/\Omega_L^2} \right)$.
Similarly, the right portion of the chain is modifed as 
\bea
    \hat{H}^{\rm eff}_{\text{half},R} = \sum_{i=N/2 + 1}^{N} \tilde{\Delta}_i\hat{\sigma}^z_i + \sum_{i=N/2 + 1}^{N} J_x \hat{\sigma}^x_{i} \hat{\sigma}^x_{i+1} + \tilde{J}_{y} \hat{\sigma}^y_{i} \hat{\sigma}^y_{i+1} + \tilde{J}_{z} \hat{\sigma}^z_{i} \hat{\sigma}^z_{i+1} - \frac{N}{2}\frac{\lambda_R^2}{\Omega_R} \hat{I} - \frac{2\lambda_R^2}{\Omega_R}\sum_{i<j = N/2 + 1}^{N} \hat{\sigma}^x_i\hat{\sigma}^x_j. 
\eea
Similarly to the left bath, we have that the spin gets suppress as $\tilde{\Delta}_i = \Delta_i e^{-\frac{2\lambda_R^2}{\Omega_R^2}}$, and the interactions via $\tilde{J}_{y(z)} = \frac{J_{y(z)}}{2} \left( 1 + e^{-8\lambda_R^2/\Omega_R^2} \right) + \frac{J_{z(y)}}{2} \left( 1 - e^{-8\lambda_R^2/\Omega_R^2} \right)$.
Lastly, the two terms at the boundary are represented, with the renormalization already taken into account as
\bea
    \hat{H}^{\rm eff}_{\text{half},\text{bound}} = J_x \hat{\sigma}^x_{N/2} \hat{\sigma}^x_{N/2 + 1} +  J_y e^{-\frac{2\lambda_{L}^2}{\Omega_{L}^2}} e^{-\frac{2\lambda_{R}^2}{\Omega_{R}^2}}\hat{\sigma}^y_{N/2} \hat{\sigma}^y_{N/2 + 1} +  J_z e^{-\frac{2\lambda_{L}^2}{\Omega_{L}^2}} e^{-\frac{2\lambda_{R}^2}{\Omega_{R}^2}}\hat{\sigma}^z_{N/2} \hat{\sigma}^z_{N/2 + 1}.
\eea
Due to the bath-induced ferromagnetic interaction, we expect to observe a transition of spin alignments in the $x$ direction from an  AFM to FM order, for spins coupled to a common bath. In contrast, the boundary spin retains their AFM interaction. This will lead to a \textit{domain wall} structure at sufficiently strong coupling. We present these transition, from an AFM order to a FM with a domain wall, in the third row of Fig. \ref{fig:heatmap} as discussed in Sec.
\ref{sec: Application to the Transverse field Ising chain}.

\subsection{Pairwise coupling model}
\label{subsec: Pairwise eff}
In this last example,  we couple each pairs of spins to a common bath. In the model,
we take
every odd site of the chain and its nearest neighbour to the right and couple them to the same bath. The total model is described by the following Hamiltonian
\begin{equation}
\label{eq:HSP}
    \hat{H}_{\text{pair}} = \hat{H}_S+\sum^{N/2}_{n=1}\hat{S}_{\text{pair},n}\sum_kt_{n,k}\left(\hat{c}^\dagger_{n,k}+\hat{c}_{n,k}\right)+\sum_{n,k}\nu_{n,k}\hat{c}^\dagger_{n,k}\hat{c}_{n,k}.
\end{equation}
Here, $\hat{S}_{\text{pair},n}=\hat{\sigma}^x_{2n-1}+\hat{\sigma}^x_{2n}$; $n\in\{1,\dots,N/2\}$ is the bath index. 
\begin{equation}
\begin{aligned}
\hat{H}^\text{eff}_{\text{pair},S}=&\sum^{N/2}_{n=1}\left(\bar{\Delta}_{2n-1}\hat{\sigma}^z_{2n-1}+\bar{\Delta}_{2n}\hat{\sigma}^z_{2n}\right)-\sum^{N/2}_{n=1}\frac{\lambda^2_n}{\Omega_n}\hat{S}^2_{\text{pair},n}+\sum_{\alpha}\sum^{N/2}_{n=1}\left(\bar{J}_{\alpha,2n-1}\hat{\sigma}^\alpha_{2n-1}\hat{\sigma}^\alpha_{2n}+\bar{J}_{\alpha,2n}\hat{\sigma}^\alpha_{2n}\hat{\sigma}^\alpha_{2n+1}\right).
\end{aligned}
\end{equation}
%
The new parameters are now defined as follows: $\Bar{\Delta}_{2n-1}=\Delta_{2n-1}\exp(-2\lambda^2_n/\Omega^2_n)$, $\bar{\Delta}_{2n}=\Delta_{2n}\exp(-2\lambda^2_n/\Omega^2_n)$, $\bar{J}_{x,2n-1}=\bar{J}_{x,2n}=J_{x}$,
\begin{equation}
    \begin{aligned}
        &\bar{J}_{y(z),2n-1}=\frac{J_{y(z)}}{2}\left(1+e^{-\frac{8\lambda^2_n}{\Omega^2_n}}\right)+\frac{J_{z(y)}}{2}\left(1-e^{-\frac{8\lambda^2_n}{\Omega^2_n}}\right),\\
        &\bar{J}_{y(z),2n} =J_{y(z)}\exp\left(-2\left(\frac{\lambda^2_{2n-1}}{\Omega^2_{2n-1}}+\frac{\lambda^2_{2n}}{\Omega^2_{2n}}\right)\right).
    \end{aligned}
\end{equation}
That is, at sufficiently strong coupling, the spin-spin interactions at the boundary of each two baths, corresponding to even site indices $(2n)$, in $y(z)$ directions will be suppressed, the result of a local-coupling. On the other hand we again generate FM interaction terms in the $x$ direction for  spins coupled to a common bath. Overall, we expect to observe an extended Neel order where the anti-ferromagnetic wavelength doubles. 
The development of this extended Neel state is presented in the fourth row of Fig. \ref{fig:heatmap} 
and is further exemplified in the main text.

\section{General Polaron Mapping and its application to the Dissipative Heisenberg chain}

\label{sec:Polaron}
In this section, we provide an alternative approach of obtaining the effective Hamiltonian via the polaron transform enacted on all modes in the bath(s). The system's Hamiltonian is again a spin chain given by
\begin{equation}
    \hat{H}_S = \sum_{i=1}^{N} \Delta_i \hat{\sigma}^z_{i} + \sum_{\alpha}\sum^{N-1}_{i=1}J_\alpha\hat{\sigma}^\alpha_i\hat{\sigma}^\alpha_{i+1}.
\end{equation}
where $\Delta_i>0$ represents the spin splitting of the $i$th qubit, $J_\alpha>0$ is the interaction strength between neighbouring spins in the $\alpha=\{x,y,z\}$ direction. We now derive the effective Hamiltonian using an alternative approach to that presented in Sec. \ref{sec: Effective Hamiltonian Mapping} for the four models considered above, fully global \ref{subsec: Fully-global pol}, fully local \ref{subsec: Fully-local pol}, half-and-half \ref{subsec: Half-half pol} and pairwise coupling \ref{subsec: Pairwise pol}.
The effective Hamiltonian created by the polaron mapping here is completely parallel to the result of the reaction-coordinate based EFFH method of Sec. \ref{sec: Application to the Dissipative Heisenberg chain}. We compare and discuss the pros and cons of these two mapping methods in Sec. \ref{sec:Disc}. 

\subsection{Fully-global coupling model}
\label{subsec: Fully-global pol}
For the fully-global coupling scheme, recall we work with the following total Hamiltonian,
\bea   
\hat{H}_\text{glob}&=&\hat{H}_S+\hat{H}_B+\hat{H}_I
\nonumber\\
&=&\hat{H}_S+\sum_k\nu_k\hat{c}^\dagger_k\hat{c}_k+
\sum^N_{n=1} \hat{\sigma}^x_n
\sum_k t_{n,k}
\left(\hat{c}^\dagger_k+\hat{c}_k\right),   
\eea
where 
here we use $n$
 as the index for the spin sites.
 Note that we can consider here a rather general case, with spins coupled at different strength to the common bath. 
 Following the procedure described in Ref.~\cite{Nick_PRB}, we perform a series of $n=1,...,N$ polaron transform via the unitary 
\begin{equation}
\label{eq: polaron transform unitary}
    \hat{W}_n = \exp(-i\hat{\sigma}^x_n\hat{B}_n/2)\quad\text{where}\quad \hat{B}_n = 2i\sum_k\frac{g_{n,k}}{\nu_k}\left(\hat{c}^\dagger_k-\hat{c}_k\right).
\end{equation}
This transformation is referred to as ``full-polaron" if the variational parameters $\{g_{n,k}\}$ are simply set to $\{t_{n,k}\}$, the original system-reservoir couplings. If, instead, the optimal values for $\{g_{n,k}\}$ are obtained by minimizing the Gibbs-Bogoliubov-Feynman upper bound on the free energy, the transformation is called ``variational" \cite{Cao_2016}. After applying consecutive $\hat{W}_n$, we subtract and add appropriate bath averaged quantities to obtain the total transformed Hamiltonian in the following form:
\begin{equation}
    \hat{H}^\text{pol} =  \hat{H}^\text{pol}_S+ \hat{H}^\text{pol}_B+ \hat{H}^\text{pol}_I.
\end{equation}
Here, the new system Hamiltonian $\hat{H}^\text{pol}_S$ is now given by
\begin{equation}
\label{eq: sys H global polaron}
\begin{aligned}
    \hat{H}^\text{pol}_S =& \sum^N_{n=1}E^{(n)}_0\hat{I}+\sum^N_{n=1}\Delta_n\langle \hat{\mathcal{C}}_n\rangle\hat{\sigma}^z_n
    +\sum^{N-1}_{n=1}\left(J_x-2E^{(n)}_I\right)\hat{\sigma}^x_{n}\hat{\sigma}^x_{n+1}
-2\sum_{n+1<m}E^{(nm)}_I\hat{\sigma}^x_n\hat{\sigma}^x_{m}\\
    %
    +&\sum^{N-1}_{n=1}\Big(\left[J_y\langle \hat{\mathcal{C}}_n\hat{\mathcal{C}}_{n+1}\rangle+J_z\langle\hat{\mathcal{S}}_n\hat{\mathcal{S}}_{n+1}\rangle\right]\hat{\sigma}^y_{n}\hat{\sigma}^y_{n+1}+\left[J_z\langle \hat{\mathcal{C}}_n\hat{\mathcal{C}}_{n+1}\rangle+J_y\langle\hat{\mathcal{S}}_n\hat{\mathcal{S}}_{n+1}\rangle\right]\hat{\sigma}^z_n\hat{\sigma}^z_{n+1}\Big)
    \end{aligned}
\end{equation}
where we define $\hat{\mathcal{C}}_n=\cos(\hat{B}_n)$ and $\hat{\mathcal{S}}_n=\sin(\hat{B}_n)$ (definition for $\hat{B}_n$ is given in Eq.~\eqref{eq: polaron transform unitary}). The averages $\langle \dots\rangle$ are done over the unchanged bath Hamiltonian $\hat{H}^\text{pol}_B = \sum_k\nu_k\hat{c}^\dagger_k \hat{c}_k$ with inverse temperature $\beta$. The renormalized system parameters now depend on the bath spectral function and temperature. Explicit forms are given as follows
\begin{equation}
\label{eq: polaron parameters}
\begin{aligned}
E^{(n)}_0 =&\sum_{n,k}\frac{g_{n,k}(g_{n,k}-2t_{n,k})}{\nu_k}\\
 E^{(n)}_I = &\sum_k\left[\left(\frac{g_{n,k}(t_{n+1,k}-g_{n+1,k}/2)}{\nu_k}\right)+\left(\frac{g_{n+1,k}(t_{n,k}-g_{n,k}/2)}{\nu_k}\right) \right]\\
    E^{(nm)}_I = &\sum_k\left[\left(\frac{g_{n,k}(t_{m,k}-g_{m,k}/2)}{\nu_k}\right)+\left(\frac{g_{m,k}(t_{n,k}-g_{n,k}/2)}{\nu_k}\right) \right]\\
    \langle\hat{\mathcal{C}}_n \rangle =&\Bigg\langle \cos\left(2i\sum_k\frac{g_{n,k}}{\nu_k}\left(\hat{c}^\dagger_k-\hat{c}_k\right)\right)\Bigg\rangle=  \exp\left(-2\sum_k\frac{g^2_{n,k}}{\nu^2_k}\coth\left(\frac{\beta\nu_k}{2}\right)\right)\\
\langle\hat{\mathcal{C}}_n\hat{\mathcal{C}}_{n+1}\rangle =
&\Bigg\langle\cos\left(2i\sum_k\frac{g_{n,k}}{\nu_k}\left(\hat{c}^\dagger_k-\hat{c}_k\right)\right)\cos\left(2i\sum_k\frac{g_{n+1,k}}{\nu_k}\left(\hat{c}^\dagger_k-\hat{c}_k\right)\right)\Bigg\rangle\\
=&\frac{1}{2}\left[\exp\left(-2\sum_k\frac{(g_{n,k}+g_{n+1,k})^2}{\nu^2_k}   \coth\left(\frac{\beta \nu_k}{2}\right)\right)+\exp\left(-2\sum_k\frac{(g_{n,k}-g_{n+1,k})^2}{\nu^2_k}   \coth\left(\frac{\beta \nu_k}{2}\right)\right)\right] \\
    \langle \hat{\mathcal{S}}_n\hat{\mathcal{S}}_{n+1}\rangle=&\Bigg\langle\sin(2i\sum_k\frac{g_{n,k}}{\nu_k}\left(\hat{c}^\dagger_k-\hat{c}_k\right))\sin\left(2i\sum_k\frac{g_{n+1,k}}{\nu_k}\left(\hat{c}^\dagger_k-\hat{c}_k\right)\right)\Bigg\rangle\\ 
    =&-\frac{1}{2}\left[\exp\left(-2\sum_k\frac{(g_{n,k}+g_{n+1,k})^2}{\nu^2_k}   \coth\left(\frac{\beta \nu_k}{2}\right)\right)-\exp\left(-2\sum_k\frac{(g_{n,k}-g_{n+1,k})^2}{\nu^2_k}   \coth\left(\frac{\beta \nu_k}{2}\right)\right)\right]\\
 \langle \hat{\mathcal{S}}_n\hat{\mathcal{C}}_{n+1}\rangle=&  
 \langle \hat{\mathcal{C}}_n\hat{\mathcal{S}}_{n+1}\rangle=0, \,\,\, 
 \langle \hat{\mathcal{S}}_n\rangle=\Bigg\langle \sin\left(2i\sum_k\frac{g_{n,k}}{\nu_k}\left(\hat{c}^\dagger_k-\hat{c}_k\right)\right)\Bigg\rangle=0. 
    \end{aligned}
\end{equation}
Note that $E^{(n)}_I$ and $E^{(nm)}_I$ are nearest-neighbour and long-ranged bath induced spin-spin interactions, respectively, that we separate in the equation for the convenience of presentation. 
In the weak coupling limit, we find
$E^{(n)}_0\rightarrow 0,\langle\hat{\mathcal{C}}_n \rangle \rightarrow 1, E^{(n)}_I\rightarrow 0, \langle\hat{\mathcal{C}}_n\hat{\mathcal{C}}_{n+1}\rangle \rightarrow 1,        E^{(nm)}_I\rightarrow 0, \langle\hat{\mathcal{S}}_n\hat{\mathcal{S}}_{n+1}\rangle\rightarrow 0 $. 
Hence, we recover the correct weak-coupling limit. 
The new system-bath interaction Hamiltonian is given by
\begin{equation}
\label{eq: global interaction Hamiltonian}
\begin{aligned}
    \hat{H}^\text{pol}_I =& \sum_n\Delta_n\Big(\left(\hat{\mathcal{C}}_n-\langle \hat{\mathcal{C}}_n\rangle\right)\hat{\sigma}^z_n-\hat{\mathcal{S}}_n\hat{\sigma}^y_n\Big)+\sum_{n,k}\hat{\sigma}^x_n\left(t_{n,k}-g_{n,k}\right)\left(\hat{b}^\dagger_k+\hat{b}_k\right)\\
    +&\sum_n\Big[J_y\left(\hat{\mathcal{C}}_n\hat{\mathcal{C}}_{n+1}-\langle\hat{\mathcal{C}}_n\hat{\mathcal{C}}_{n+1}\rangle\right)+J_z\left(\hat{\mathcal{S}}_n\hat{\mathcal{S}}_{n+1}-\langle \hat{\mathcal{S}}_n\hat{\mathcal{S}}_{n+1}\rangle \right)\Big]\hat{\sigma}^y_n\hat{\sigma}^y_{n+1}\\
    +&\sum_j\Big[J_z\left(\hat{\mathcal{C}}_n\hat{\mathcal{C}}_{n+1}-\langle\hat{\mathcal{C}}_n\hat{\mathcal{C}}_{n+1}\rangle\right)+J_y\left(\hat{\mathcal{S}}_n\hat{\mathcal{S}}_{n+1}-\langle \hat{\mathcal{S}}_n\hat{\mathcal{S}}_{n+1}\rangle \right)\Big]\hat{\sigma}^z_n\hat{\sigma}^z_{n+1}\\
    +&\sum_n\Big[\left(J_y\hat{\mathcal{C}}_n\hat{\mathcal{S}}_{n+1}-J_z\hat{\mathcal{S}}_n\hat{\mathcal{C}}_{n+1}\right)\hat{\sigma}^y_n\hat{\sigma}^z_{n+1}+\left(J_y\hat{\mathcal{S}}_n\hat{\mathcal{C}}_{n+1}-J_z\hat{\mathcal{C}}_n\hat{\mathcal{S}}_{n+1}\right)\hat{\sigma}^z_n\hat{\sigma}^y_{n+1}\Big].
    \end{aligned}
\end{equation}
Here, the bath averaged quantities are identical to the expressions summarized in Eq.~\eqref{eq: polaron parameters}.
If we now consider a symmetric model, with the full-polaron mapping ($J_y=J_z$ and $g_{n,k}=t_{n,k}=t_k$), the interaction Hamiltonian reduces to 
\begin{equation}
    \hat{H}^\text{pol}_I = \sum_n\Delta_n\Big(\left(\hat{\mathcal{C}}_n-\langle \hat{\mathcal{C}}_n\rangle\right)\hat{\sigma}^z_n-\hat{\mathcal{S}}_n\hat{\sigma}^y_n\Big),
\end{equation}
which will be indeed weak when $\Delta_n$ is small. As a result, we can safely approximate the equilibrium steady-state of the dissipative chain to be
\begin{equation}
    \hat{\rho}_S^{\text{pol}}=
    e^{-\beta\hat{H}^{\text{pol}}_S}
    /Z^{\text{pol}}_S,
\end{equation}
from which various steady-state observable can be computed. $Z^{\text{pol}}_S$ is the partition function of the model.

Next, we express the coefficients in the Hamiltonian using the continuous spectral density function, $K_n(\omega)= \sum_k t^2_{n,k}\delta(\omega-\nu_k)$.
The full-polaron mapping is known to be most accurate for the super-Ohmic spectral function given by 
\bea
K_n(\omega)= \alpha_n\frac{\omega^3}{\omega^2_c}e^{-\omega/\omega_c},
\eea
where $\alpha_n$ is a dimensionless coupling strength for n$^\text{th}$ spin and $\omega_c$ is the cutoff (or characteristic) frequency of the bath. Using this model, we are able to represent all the parameters in Eq.~\eqref{eq: polaron parameters} in terms these parameters, and the bath's temperature. Specifically,
\begin{equation}
\begin{aligned}
    E^{(n)}_0 &= -2\omega_c\sum_n\alpha_n\\
    E^{(n)}_I & = 2\omega_c\sqrt{\alpha_n\alpha_{n+1}}\\
    E^{(nm)}_I& = 2\omega_c\sqrt{\alpha_n\alpha_{m}}\\
    \langle\hat{\mathcal{C}}_n\rangle &= \exp\left(-2\int^\infty_0 d\omega\frac{K_n(\omega)}{\omega^2}\coth\left(\frac{\beta \omega}{2}\right)\right)\\ 
\langle\hat{\mathcal{C}}_n\hat{\mathcal{C}}_{n+1}\rangle 
    &=\frac{1}{2}\left[\exp\left(-2\int d\omega \frac{K_n(\omega)+2\sqrt{K_n(\omega)K_{n+1}(\omega)}+K_{n+1}(\omega)}{\omega^2}\coth\left(\frac{\beta\omega}{2}\right)\right)\right]\\
    &+\frac{1}{2}\left[\exp\left(-2\int d\omega \frac{K_n(\omega)-2\sqrt{K_n(\omega)K_{n+1}(\omega)}+K_{n+1}(\omega)}{\omega^2}\coth\left(\frac{\beta\omega}{2}\right)\right)\right]\\
     \langle \hat{\mathcal{S}}_n\hat{\mathcal{S}}_{n+1}\rangle 
      =&-\frac{1}{2}\left[\exp\left(-2\int d\omega \frac{K_n(\omega)+2\sqrt{K_n(\omega)K_{n+1}(\omega)}+K_{n+1}(\omega)}{\omega^2}\coth\left(\frac{\beta\omega}{2}\right)\right)\right]\\
    &+\frac{1}{2}\left[\exp\left(-2\int d\omega \frac{K_n(\omega)-2\sqrt{K_n(\omega)K_{n+1}(\omega)}+K_{n+1}(\omega)}{\omega^2}\coth\left(\frac{\beta\omega}{2}\right)\right)\right].\\
    \end{aligned}
\end{equation}
In the special case when all the spins couple to the bath in an identical manner,
\bea 
E_I=2\omega_c\alpha
\label{eq:EI}
\eea 
is identified as the bath-induced spin-spin interaction energy (with a factor of 2 further appearing in Eq.~\eqref{eq: sys H global polaron}).
As expected, this interaction energy depends on properties of the bath: Its characteristic frequency and its coupling parameter to the system. Interestingly, it does not depend on the temperature of the bath; temperature dependence of $E_I$ could show up in the variational polaron treatment \cite{Nick_PRB} and when considering other types of baths, e.g., spin baths. 

\subsection{Fully local coupling model}
\label{subsec: Fully-local pol}

We now study the Heisenberg model with a  fully-local coupling scheme, that is, with individual separate baths coupled to each site. The total Hamiltonian is
\begin{equation}
    \hat{H}_\text{loca} = \hat{H}_S+\sum^N_{n=1}\hat{S}_\text{loca,n}\sum_kt_{n,k}\left(\hat{c}^\dagger_{n,k}+\hat{c}_{n,k}\right)+\sum_{n,k}\nu_{n,k}\hat{c}^\dagger_{n,k}\hat{c}_{n,k}
\end{equation}
where $\hat{S}_{\text{loca},n}=\hat{\sigma}^x_n$. After consecutive polaron transformation  via the unitary 
\begin{equation}
    \hat{W}_n = \exp(-i\hat{\sigma}^x_n\hat{B}_n/2)\quad\text{where}\quad \hat{B}_n = 2i\sum_k\frac{g_{n,k}}{\nu_k}\left(\hat{c}^\dagger_{n,k}-\hat{c}_{n,k}\right).
\end{equation}
we obtain the polaron-transformed Hamiltonian
\begin{equation}
    \hat{H}^\text{pol} = \hat{H}^\text{pol}_S+\hat{H}^\text{pol}_B+\hat{H}^\text{pol}_I,
\end{equation}
where the system's Hamiltonian is
\begin{equation}
    \hat{H}^\text{pol}_S = \sum^{N}_{n=1}\Delta_n\langle \hat{\mathcal{C}}_n\rangle+\sum^{N-1}_{n=1}\left(J_x\hat{\sigma}^x_n\hat{\sigma}^x_{n+1}+J_y\langle \hat{\mathcal{C}}_n\rangle\langle \hat{\mathcal{C}}_{n+1}\rangle\hat{\sigma}^y_n\hat{\sigma}^y_{n+1}+J_z\langle \hat{\mathcal{C}}_n\rangle\langle \hat{\mathcal{C}}_{n+1}\rangle\hat{\sigma}^z_n\hat{\sigma}^z_{n+1}\right).
    \label{eq:HSpolL}
\end{equation}
Recall that averages are done with respect to the state of the baths, which are given, as before, by
\begin{equation}
    \hat{H}^\text{pol}_B = \sum_{n,k}\nu_{n,k}\hat{c}^\dagger_{n,k}\hat{c}_{n,k},
\end{equation}
and the system-bath interaction Hamiltonian is
\begin{equation}
\label{eq: local interaction Hamiltonian}
\begin{aligned}
    \hat{H}^\text{pol}_I = &\sum^N_{n=1}\Delta_n\left(\left[\hat{\mathcal{C}}_n-\langle\hat{\mathcal{C}}_n\rangle\right]\hat{\sigma}^z_n-\hat{\mathcal{S}}_n\hat{\sigma}^y_n\right)\\
    %
    %
&\sum^{N-1}_{n=1}\left[J_y\hat{\mathcal{C}}_n \hat{\mathcal{C}}_{n+1}
-J_y\langle\hat{\mathcal{C}}_n \rangle\langle \hat{\mathcal{C}}_{n+1}\rangle
+J_z\hat{\mathcal{S}}_n\hat{\mathcal{S}}_{n+1}\right]\hat{\sigma}^y_n\hat{\sigma}^y_{n+1}\\
%
    +&\sum^{N-1}_{n=1}\left[J_z\hat{\mathcal{C}}_n\hat{\mathcal{C}}_{n+1}-J_z\langle\hat{\mathcal{C}}_n \rangle\langle \hat{\mathcal{C}}_{n+1}\rangle+J_y\hat{\mathcal{S}}_n\hat{\mathcal{S}}_{n+1}\right]\hat{\sigma}^z_n\hat{\sigma}^z_{n+1}\\
    +&\sum^{N-1}_{n=1}\Big(\left[J_y\hat{\mathcal{C}}_n\hat{\mathcal{S}}_{n+1}-J_z\hat{\mathcal{S}}_n\hat{\mathcal{C}}_{n+1}\right]\hat{\sigma}^y_n\hat{\sigma}^z_{n+1}+\left[J_y\hat{\mathcal{S}}_n\hat{\mathcal{C}}_{n+1}-J_z\hat{\mathcal{S}}_n\hat{\mathcal{C}}_{n+1}\right]\hat{\sigma}^z_n\hat{\sigma}^y_{n+1}\Big).
    \end{aligned}
\end{equation}
The expressions in this Hamiltonian are given by  Eq.~\eqref{eq: polaron parameters}, before enacting the thermal average. The important difference is that each site $n$ here couples to a distinct bath, and the baths are uncorrelated, leading to 
$\langle \hat{\mathcal{S}}_n\hat{\mathcal{S}}_{n+1} \rangle =0$
and
$\langle \hat{\mathcal{C}}_n\hat{\mathcal{C}}_{n+1} \rangle = \langle \hat{\mathcal{C}}_n\rangle\langle\hat{\mathcal{C}}_{n+1} \rangle$.

\subsection{Half-half coupling model}
\label{subsec: Half-half pol}

Next, we move on to the half-and-half coupling scheme. Here, an $N$-site chain is coupled to two reservoirs, $L$ and $R$, according to the following Hamiltonian
\begin{equation}
    \hat{H}_\text{half} = \hat{H}_S+\sum_n\hat{S}_\text{half,n}\sum_kt_{n,k}\left(\hat{c}^\dagger_{n,k}+\hat{c}_{n,k}\right)+\sum_{n,k}\nu_{n,k}\hat{c}^\dagger_{n,k}\hat{c}_{n,k}.
\end{equation}
The system $\hat{H}_S$ is again given by Eq.~\eqref{eq: Heisenberg chain}, $n=\{L,R\}$.
As for the coupling operators, we use
$\hat{S}_\text{half,L}=\sum^{N/2}_{n=1}\hat{\sigma}^x_n$, and $\hat{S}_{\text{half},R}=\sum^N_{N/2+1}\hat{\sigma}^x_n$.
That is, the left half of the chain is coupled to bath $L$ while the right half is coupled to bath $R$. We can already infer the effect of the half-and-half coupling scheme. That is, the system develops a long-range spin-spin interaction in the $x$ direction for all spins coupled to a common bath. 
On the other hand, the $J_{y(z)}$ interaction at the boundary between the two segments is expected to be suppressed. To observe the effect of the transformation more clearly, we  break $\hat{H}_S$ into three sectors,
\begin{equation}
\label{eq:HSpolH}
\begin{aligned}
\hat{H}_S = \hat{H}_{S,L}+\hat{H}_{S,\text{boundary}}+\hat{H}_{S,R}.
\end{aligned}
\end{equation}
Here,
\begin{equation}
    \begin{aligned}
        \hat{H}_{S,L} =& \sum^{N/2}_{n=1}\Delta_n\hat{\sigma}^z_n+\sum_{\alpha}\sum^{N/2-1}_{n=1}J_\alpha\hat{\sigma}^\alpha_n\hat{\sigma}^\alpha_{n+1}\\
        \hat{H}_{S,\text{boundary}} =& \sum_{\alpha}J_\alpha \hat{\sigma}^\alpha_{N/2}\hat{\sigma}^\alpha_{N/2+1}\\
        \hat{H}_{S,R} =& \sum^{N}_{n=N/2+1}\Delta_n\hat{\sigma}^z_n+\sum_{\alpha}\sum^{N-1}_{n=N/2+1}J_\alpha\hat{\sigma}^\alpha_n\hat{\sigma}^\alpha_{n+1}
    \end{aligned}
\end{equation}
Note that the $\hat{H}_{S,L}$ ($\hat{H}_{S,R}$) is the left (right) half of the chain that are coupled to a common bath $L$ ($R$). The boundary term only involves the spin-spin interaction at the right most end of chain $L$ and the left-most end of chain $R$. The polaron transformed Hamiltonian is
\begin{equation}
    \hat{H}^\text{pol}_S= \hat{H}^\text{pol}_{S,L}+\hat{H}^\text{pol}_{S,\text{boundary}}+\hat{H}^\text{pol}_{S,R}
\end{equation}
where
\begin{equation}
\begin{aligned}
    \hat{H}^\text{pol}_{S,L} =& \sum^{N/2}_{n=1}E^{(n)}_0\hat{I}+\sum^{N/2}_{n=1}\Delta_n\langle \hat{\mathcal{C}}_n\rangle\hat{\sigma}^z_n+\sum^{N/2-1}_{n=1}\left(J_x-2E^{(n)}_I\right)\hat{\sigma}^x_{n}\hat{\sigma}^x_{n+1}-\sum_{|n-m|>1}E^{(nm)}_I\hat{\sigma}^x_n\hat{\sigma}^x_m\\
    +&\sum^{N/2-1}_{n=1}\Big(\left[J_y\langle \hat{\mathcal{C}}_n\hat{\mathcal{C}}_{n+1}\rangle+J_z\langle\hat{\mathcal{S}}_n\hat{\mathcal{S}}_{n+1}\rangle\right]\hat{\sigma}^y_{n}\hat{\sigma}^y_{n+1}+\left[J_z\langle \hat{\mathcal{C}}_n\hat{\mathcal{C}}_{n+1}\rangle+J_y\langle\hat{\mathcal{S}}_n\hat{\mathcal{S}}_{n+1}\rangle\right]\hat{\sigma}^z_n\hat{\sigma}^z_{n+1}\Big),
    \end{aligned}
\end{equation}
\begin{equation}
\begin{aligned}
    \hat{H}^\text{pol}_{S,R} =& \sum^{N}_{n=N/2+1}E^{(n)}_0\hat{I}+\sum^{N}_{n=N/2+1}\Delta_n\langle \hat{\mathcal{C}}_n\rangle\hat{\sigma}^z_n+\sum^{N-1}_{n=N/2+1}\left(J_x-2E^{(n)}_I\right)\hat{\sigma}^x_{n}\hat{\sigma}^x_{n+1}-\sum_{|n-m|>1}E^{(nm)}_I\hat{\sigma}^x_n\hat{\sigma}^x_m\\
    +&\sum^{N-1}_{n=N/2+1}\Big(\left[J_y\langle \hat{\mathcal{C}}_n\hat{\mathcal{C}}_{n+1}\rangle+J_z\langle\hat{\mathcal{S}}_n\hat{\mathcal{S}}_{n+1}\rangle\right]\hat{\sigma}^y_{n}\hat{\sigma}^y_{n+1}+\left[J_z\langle \hat{\mathcal{C}}_n\hat{\mathcal{C}}_{n+1}\rangle+J_y\langle\hat{\mathcal{S}}_n\hat{\mathcal{S}}_{n+1}\rangle\right]\hat{\sigma}^z_n\hat{\sigma}^z_{n+1}\Big),
    \end{aligned}
\end{equation}
and the boundary term is
\begin{equation}
\hat{H}^\text{pol}_{S,\text{bound}} = J_x\hat{\sigma}^x_{N/2}\hat{\sigma}^x_{N/2+1}+J_y\langle \hat{\mathcal{C}}_{N/2}\rangle\langle \hat{\mathcal{C}}_{N/2+1}\rangle\hat{\sigma}^y_{N/2}\hat{\sigma}^y_{N/2+1}+J_z\langle \hat{\mathcal{C}}_{N/2}\rangle\langle \hat{\mathcal{C}}_{N/2+1}\rangle\hat{\sigma}^z_{N/2}\hat{\sigma}^z_{N/2+1}
\end{equation}
Note that at strong coupling, the $L$ and $R$ segments of the chain interact with each other via the $\sigma^x_{N/2}\sigma^x_{N/2+1}$ coupling at the boundary thus preferring the anti-ferromagnetic exchange $J_x>0$. In contrast, the rest of the spins will aligned themselves in their respective baths in the $x$ direction due to the FM term over dominating at strong coupling.

Expressions for the renormalized parameters are given by Eq.~\eqref{eq: polaron parameters}. The system-bath interaction Hamiltonian can be trivially obtained by combining Eq.~\eqref{eq: global interaction Hamiltonian} and Eq.~\eqref{eq: local interaction Hamiltonian}. The bath Hamiltonians remain unaffected by the polaron transform. 

\subsection{Pairwise coupling model}
\label{subsec: Pairwise pol}

Finally, we examine the pairwise coupling scheme given by the following Hamiltonian
\begin{equation}
    \hat{H}_{\text{pair}} = \hat{H}_S+\sum^{N/2}_{n=1}\hat{S}_\text{pair,n}\sum_kt_{n,k}\left(\hat{c}^\dagger_{n,k}+\hat{c}_{n,k}\right)+\sum_{n,k}\nu_{n,k}\hat{c}^\dagger_{n,k}\hat{c}_{n,k}
\end{equation}
where $\hat{S}_{\text{pair},n}=\hat{\sigma}^x_{2n-1}+\hat{\sigma}^x_{2n}$; $n\in\{1,\dots,N/2\}$ is the bath index.
Here, every odd spin will be coupled to a common bath 
together with the even numbered spin to its right.
The resulting system's Hamiltonian after the polaron transformation is
\begin{equation}
\begin{aligned}
\hat{H}^\text{pol}_{\text{pair},S}=&\sum^{N/2}_{n=1}\left(\Delta_{2n-1}\langle \hat{\mathcal{C}}_{2n-1}\rangle\hat{\sigma}^z_{2n-1}+\Delta_{2n}\langle\hat{\mathcal{C}}_{2n}\rangle\hat{\sigma}^z_{2n}\right)\\
+&\sum^{N/2}_{n=1}J_x\hat{\sigma}^x_{2n}\hat{\sigma}^x_{2n+1}+\sum^{N/2}_{n=1}\left(J_x-2E^{(2n-1,2n)}_I\right)\hat{\sigma}^x_{2n-1}\hat{\sigma}^x_{2n}\\
+&\sum^{N/2}_{n=1}\left[\left[J_y\langle \hat{\mathcal{C}}_{2n-1}\hat{\mathcal{C}}_{2n}\rangle+J_z\langle\hat{\mathcal{S}}_{2n-1}\hat{\mathcal{S}}_{2n}\rangle\right]\hat{\sigma}^y_{2n-1}\hat{\sigma}^y_{2n}+\left[J_z\langle \hat{\mathcal{C}}_{2n-1}\hat{\mathcal{C}}_{2n}\rangle+J_y\langle\hat{\mathcal{S}}_{2n-1}\hat{\mathcal{S}}_{2n}\rangle\right]\hat{\sigma}^z_{2n-1}\hat{\sigma}^z_{2n}\right]\\
+&\sum^{N/2-1}_{n=1}\left[J_y\langle \hat{\mathcal{C}}_{2n}\rangle\langle \hat{\mathcal{C}}_{2n+1}\rangle\hat{\sigma}^y_{2n}\hat{\sigma}^y_{2n+1}+J_z\langle \hat{\mathcal{C}}_{2n}\rangle\langle \hat{\mathcal{C}}_{2n+1}\rangle\hat{\sigma}^z_{2n}\hat{\sigma}^z_{2n+1}\right].\\
\end{aligned}
\label{eq:HSpolP}
\end{equation}
%
%
The new system-bath interaction Hamiltonian can be obtained by combining Eq.~\eqref{eq: global interaction Hamiltonian}  and Eq.~\eqref{eq: local interaction Hamiltonian}. 

\section{Discussion of the two mapping approaches}
\label{sec:Disc}

We summarize the two mapping approaches.
In the EFFH method, we first extract a collective reaction coordinate mode, polaron transform this specific mode and truncate it. This approach was exercised on spin chains in Sec.~\ref{sec: Application to the Dissipative Heisenberg chain}.
A polaron approach on the full bath was detailed in Sec.~\ref{sec:Polaron}.
Notably, the two mapping approaches provide the {\it same} form for the system's Hamiltonian.
This can be clearly observed for, e.g., the global coupling scheme by comparing the system terms in Eq.~\eqref{eq:h_eff_tilted_heisenberg}
to the system's Hamiltonian in Eq.~\eqref{eq: sys H global polaron}.
Similarly, in the local-bath scheme we get Eq.~\eqref{eq:H2Qeff} corresponding to Eq.~\eqref{eq:HSpolL}.
In the half-and-half coupling model, Eq.~\eqref{eq:HSH} matches Eq.~\eqref{eq:HSpolH}.
In the pairwise coupling model,
Eq.~\eqref{eq:HSP} corresponds to Eq.~\eqref{eq:HSpolP}.
The significance of this agreement is that we predict that equilibrium expectation values of these system's Hamiltonian will display the same physics. Moreover, the prediction on the expected magnetic order hold for any spectral function---so long as one can confirm that in the EFFH model, the residual coupling is weak, and that the polaron approach provides a good approximation to exact results.
Here, we tested the EFFH model assuming a Brownian spectral function for the bath, and the polaron mapping assuming a super-Ohmic function.

As for the system-bath interaction Hamiltonian, the 
EFFH and polaron mapping methods yield different terms. Most notably, the interaction Hamiltonian under the polaron mapping are more complex bringing more terms.
Whether relaxation dynamics to equilibrium or steady state will be noticeably different in this two methods remains a topic of future studies. 

The advantages of the EFFH method is the simplicity of derivation compared to the polaron method. However, the polaron approach here was general in the sense that different spins could couple with different strength to the same bath. In contrast, the EFFH mapping needs to be further extended to handle such an effect. Here the difficulty would lie in building a RC that balances different coupling strength to the different spins.


\section{Application: The Transverse field Ising chain} 
\label{sec: Application to the Transverse field Ising chain}

In this section, we study the magnetic order in the Ising model. First, we consider the case of an Ising chain coupled to a  global bath, then look at other coupling schemes.

We turn the general Heisenberg model into an Ising chain by setting $J_y=J_z=0$. This allows us to clearly analyze and visualize the preferred magnetic order under system-bath coupling.
The starting Hamiltonian is 
\begin{equation}
\hat{H}^\text{Ising}_S=\sum^N_{i=1}\Delta_i\hat{\sigma}^z_i+\sum^{N-1}_{i=1}J_x\hat{\sigma}^x_i\hat{\sigma}^x_{i+1}.
\end{equation}
%
In the global-coupling scheme, the effective system's Hamiltonian, obtained from either the EFFH treatment or the polaron approach, is given by 
\begin{equation}
\begin{aligned}
\hat{H}^\text{eff}_{\text{glob},S} =&\sum^N_{i=1}\tilde{\Delta}_i\hat{\sigma}^z_i+\sum^{N-1}_{i=1}\tilde{J}_x\hat{\sigma}^x_i\hat{\sigma}^x_{i+1}-\frac{\lambda^2}{\Omega}\hat{S}^2_\text{glob}\\
\hat{H}^\text{pol}_{\text{glob},S}=&\sum^N_{n=1}E^{(n)}_0\hat{I}+\sum^N_{n=1}\Delta_n\langle \hat{\mathcal{C}}_n\rangle\hat{\sigma}^z_n+\sum^{N-1}_{n=1}\left(J_x-2E^{(n)}_I\right)\hat{\sigma}^x_{n}\hat{\sigma}^x_{n+1}-2\sum_{n+1<m}E^{(nm)}_I\hat{\sigma}^x_n\hat{\sigma}^x_m.
\end{aligned}
\end{equation}
It is obtained by  setting $J_y=J_z=0$ in Eq.~\eqref{eq:h_eff_tilted_heisenberg} and Eq.~\eqref{eq: sys H global polaron}. The most important aspect of this model is the development of bath-induced long range spin-spin coupling with magnitude
$\frac{2\lambda^2}{\Omega}$ in the language of the EFFH mapping and $4\omega_c \alpha$ in the polaron super-Ohmic bath mapping case, see Eq. (\ref{eq:EI}).

\begin{figure}[htpb]
\fontsize{13}{10}\selectfont 
\centering
\includegraphics[width=0.7\columnwidth]{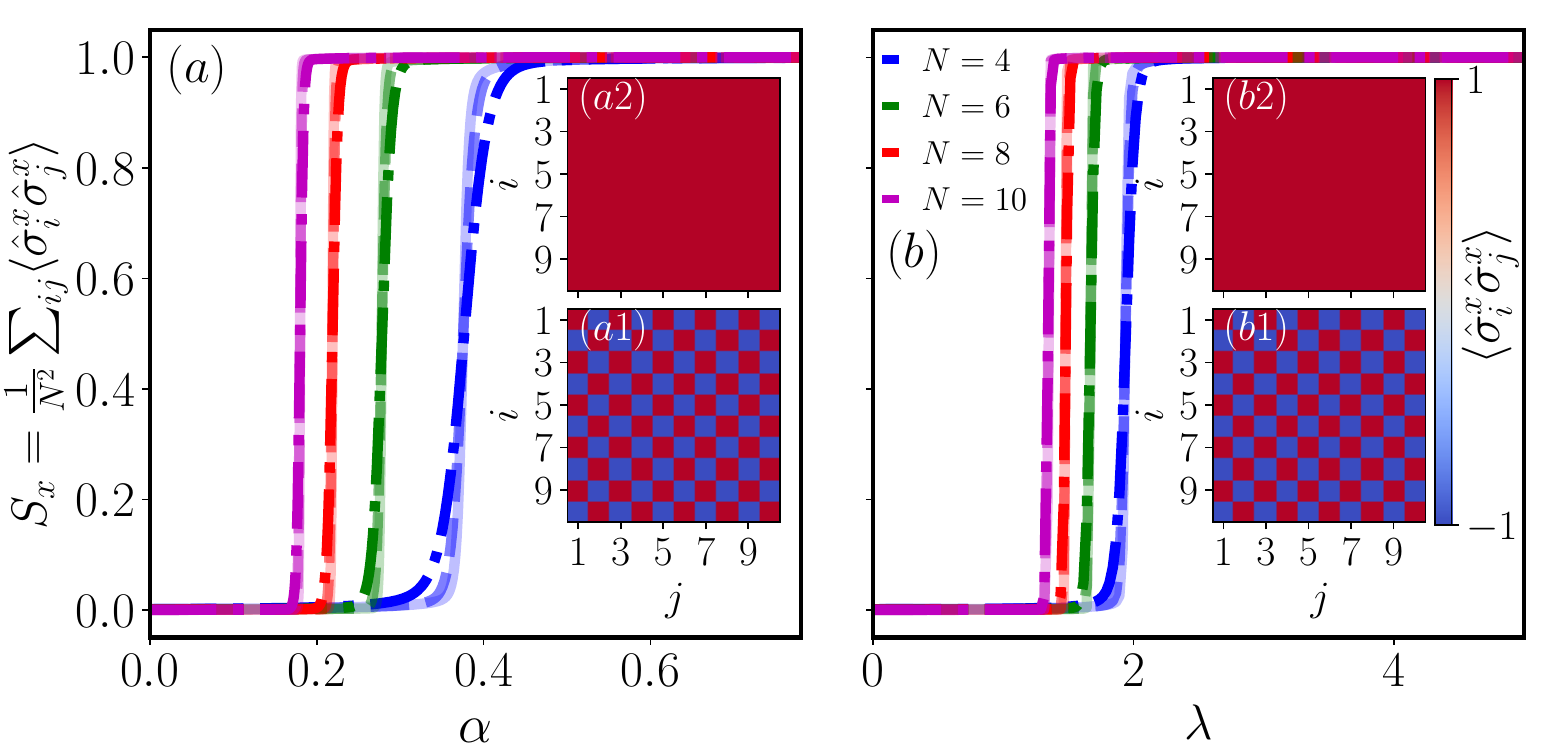}
\caption{Structure factor $S_x=\frac{1}{N^2}\sum_{ij}\langle \hat{\sigma}^x_i\hat{\sigma}^x_j\rangle$ for a globally-coupled transverse field Ising chain.
We consider models corresponding to (a) super-Ohmic and (b) Brownian baths. The structure factor indicates a clear crossover from an AFM to FM order due to the coupling to the bath with results displayed for different chain lengths, $N=4,6,8,10$ sites, and temperatures; dark to light corresponding to $T=0.2,0.1,0.05$.
 Insets $(a_1)$ and $(a_2)$ correspond to spin correlations, $\langle\hat{\sigma}^x_i\hat{\sigma}^x_j\rangle$, at $\alpha=0$ and $\alpha=0.3$, respectively for $N=10$ with $T=\Delta=0.1$ and $J_x=1$.
Insets $(b_1)$ and $(b_2)$ correspond to $\langle\hat{\sigma}^x_i\hat{\sigma}^x_j\rangle$ at $\lambda=0$ 
and $\lambda=3$ 
respectively for $N=10$ with $T=\Delta=0.1$ and $J_x=1$. 
}
\label{fig:2_sup}
\end{figure}
In Figure~\ref{fig:2_sup}, we plot the structure factor $S_x=\frac{1}{N^2}\sum_{ij}\langle\hat{\sigma}^x_i\hat{\sigma}^x_j\rangle$ 
as a function of coupling strength to the bath.
The nature of the bath affects the renormalization of parameters and the bath-generated spin coupling $E_I$.
As a demonstration, we assume a super-Ohmic bath spectral function in panel (a) of Figure~\ref{fig:2_sup}.
Panel (b) of that figure could correspond to any spectral function, as long as the residual coupling is weak. Here we associate it as an example with a Brownian spectral function expressed in Eq.~\eqref{eq:Brown}.
Regardless of the bath spectral density, the structure function evinces that the preferred magnetic order in the Ising chain with the original AFM interaction goes into an FM order as the coupling to the bath increases. 
Results are displayed for different chain sizes and at different temperatures. Furthermore, we display as a heat map spin correlations in the $x$ direction at both weak and strong coupling, exemplifying the AFM and the FM order, respectively. 


\begin{figure}[htbp]
\fontsize{6}{10}\selectfont 
\centering
\includegraphics[width=1\columnwidth]{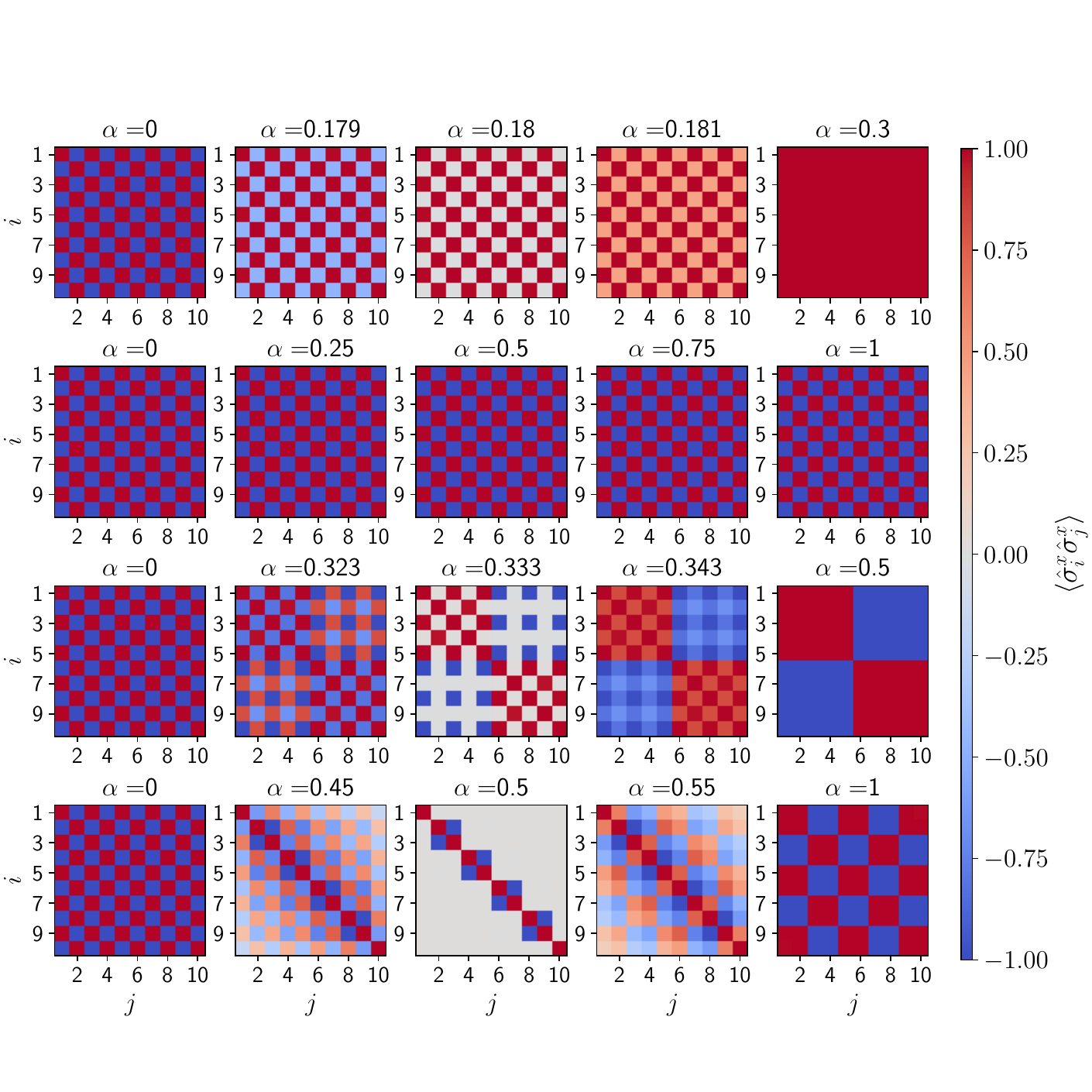}
\caption{Spin-spin correlations in the $x$ direction for the transverse field Ising chain coupled to super-Ohmic bath(s). Rows represents fully-global, fully-local, half-and-half, and pairwise coupling schemes from top to bottom corresponding to Fig. 1 in the Main text. Columns represent five different coupled strengths to clearly demonstrate the crossover of the preferred magnetic alignment for each coupling schemes (with the specific coupling strengths indicated in the title). Parameters are chosen to be $N=10$, $J_x=1$, $T=\Delta=0.1$. 
}
\label{fig:heatmap}
\end{figure}

Finally, in Figure~\ref{fig:heatmap}, we 
present a comprehensive picture of bath-induced magnetic order in the transverse field Ising chain with different schemes for the bath coupling.
We present the spin-spin correlations, $\langle \hat{\sigma}_i\hat{\sigma}_j\rangle$, and use parameters corresponding to a super-Ohmic bath spectral density function.

We study the four models of Fig.~1 (Main) and present the spin-spin correlations in the $x$ direction in Fig.~\ref{fig:heatmap}.
The first row corresponds to the global-bath model. Here, the system shows the transition from an AFM to a FM order with increasing coupling $\alpha$.
The second row corresponds to the local bath case. Here, the interaction with the bath suppresses the energy splitting thus turning the model into the zero-field Ising model.
The third row shows results for the half-and-half coupling model.
Here, we observe the development of a domain wall between the two segments, each showing internally an FM order. The last row displays the development of an extended Neel phase in the pairwise coupling model. Here each pair of spins align in the same direction, but the interaction between every pair is still anti-ferromagnetic. 
This case is described in details in Fig. 3 of the Main text.


\bibliographystyle{apsrev4-1}
\bibliography{bibliography}
\twocolumngrid